\documentclass[journal]{IEEEtran}
\usepackage{setspace}
\usepackage{amssymb}
\usepackage{amsfonts}
\usepackage{eucal}
\usepackage{bbm}
\usepackage{amsfonts}
\usepackage{graphicx}
\usepackage{amsmath}
\usepackage{fancyhdr}
\usepackage{stfloats}
\usepackage{multirow}
\usepackage{algorithm}
\usepackage{algorithmicx}
\usepackage{cite}
\usepackage[bookmarks=false]{}
\usepackage{amsthm}
\usepackage{algorithm}
\usepackage{mathbbol}
\usepackage{amsfonts}
\usepackage{graphicx}
\usepackage{amsmath}
\usepackage{amssymb}
\usepackage{latexsym}
\usepackage{graphicx}
\usepackage{cite}
\usepackage{url}
\usepackage{stfloats}
\usepackage{mathrsfs}
\usepackage{setspace}
\usepackage{booktabs}
\usepackage {latexsym}
\usepackage {graphicx}
\usepackage {cite}
\usepackage {url}
\usepackage {stfloats}
\usepackage {mathrsfs}
\usepackage{amssymb}
\usepackage{amsfonts}
\usepackage{amsmath}
\usepackage{amsthm}
\usepackage{mathbbol}
\usepackage{algorithm}
\usepackage{amssymb}
\usepackage{bbm}
\usepackage{booktabs}
\usepackage{cite}
\usepackage{fancyhdr}
\usepackage{graphicx}
\usepackage{latexsym}
\usepackage{multirow}
\usepackage{mathrsfs}
\usepackage{setspace}
\usepackage{stfloats}
\usepackage[tight,footnotesize]{subfigure}
\usepackage{url}
\usepackage {latexsym}
\usepackage{pifont}

\IEEEoverridecommandlockouts
\usepackage[table]{xcolor}
\usepackage{cite}
\usepackage{amsmath,amssymb,amsfonts}
\usepackage{cuted}
\usepackage{graphicx}
\usepackage{textcomp}
\usepackage{xcolor}
\usepackage{verbatim}
\usepackage{float}
\usepackage{bm}
\usepackage{geometry}
\usepackage{setspace}
\usepackage{stfloats}
\usepackage{bigstrut,multirow,rotating}

\usepackage{algorithm}
\usepackage{algorithmicx}
\usepackage{algpseudocode}
\usepackage{amsthm}
\usepackage{color}
\usepackage{dsfont}
\usepackage{bbm, dsfont}
\usepackage[tight,footnotesize]{subfigure}

\def\BibTeX{{\rm B\kern-.05em{\sc i\kern-.025em b}\kern-.08em
    T\kern-.1667em\lower.7ex\hbox{E}\kern-.125emX}}

\theoremstyle{plain}

\ifCLASSINFOpdf
\else
\fi
\makeatletter
\renewcommand{\maketag@@@}[1]{\hbox{\m@th\normalsize\normalfont#1}}%
\makeatother

\DeclareMathAlphabet{\mathcal}{OMS}{cmsy}{m}{n}
\let\mathbb\relax 
\DeclareMathAlphabet{\mathbb}{U}{msb}{m}{n}

\usepackage{threeparttable}

\begin{document}
\clearpage
\title{\huge \textbf{A Survey on Indoor Visible Light Positioning Systems: Fundamentals, Applications, and Challenges}}
%
\author{Zhiyu Zhu, Yang Yang, ~\IEEEmembership{Member,~IEEE},  Mingzhe Chen,  ~\IEEEmembership{Senior Member,~IEEE}, Caili Guo, ~\IEEEmembership{Senior Member,~IEEE}, Julian Cheng, ~\IEEEmembership{Fellow,~IEEE}, and Shuguang Cui, ~\IEEEmembership{Fellow,~IEEE}
\thanks{This work was supported in part by the National Natural Science Foundation of China  under Grant 62371065, Grant 61901047, and Grant 61871047, and in part by the Fundamental Research Funds for the Central Universities under Grant 2021XD-A01-1.}
\thanks{Z. Zhu and Y. Yang are with the Beijing Key Laboratory of Network System Architecture and Convergence, School of Information and Communication Engineering, Beijing University of Posts and Telecommunications, Beijing 100876, China (e-mail: zhuzy@bupt.edu.cn; yangyang01@bupt.edu.cn).
}
\thanks{M. Chen is with the Department of Electrical and Computer Engineering and Institute for Data Science and Computing, University of Miami, Coral Gables, FL, 33146, USA (e-mail: mingzhe.chen@miami.edu).
}
\thanks{C. Guo is with the Beijing Laboratory of Advanced Information Networks, School of Information and Communication Engineering, Beijing University of Posts and Telecommunications, Beijing 100876, China (e-mail: guocaili@bupt.edu.cn).
}
\thanks{J. Cheng is with the Faculty of Applied Science, School of Engineering,
The University of British Columbia, Kelowna, BC V1V 1V7, Canada (e-mail:
julian.cheng@ubc.ca).
}
\thanks{S. Cui is currently with the School of Science and Engineering, Shenzhen Research Institute of Big Data and Future Network of Intelligence Institute (FNii), the Chinese University of Hong Kong, Shenzhen, China, 518172 (email: shuguangcui@cuhk.edu.cn).
}
}
%
%
%
%
\maketitle
\thispagestyle{empty}
\vspace{0cm}
\begin{abstract}
The growing demand for location-based services in areas like virtual reality, robot control, and navigation has intensified the focus on indoor localization.
Visible light positioning (VLP), leveraging visible light communications (VLC), becomes a promising indoor positioning technology due to its high accuracy and low cost. This paper provides a comprehensive survey of VLP systems. In particular, since VLC lays the foundation for VLP, we first present a detailed overview of the principles of VLC. 
Then, we provide an in-depth overview of VLP algorithms. 
The performance of each positioning algorithm is also compared in terms of various metrics such as accuracy, coverage, and orientation limitation. Beyond the physical layer studies, the network design for a VLP system is also investigated, including multi-access technologies, resource allocation, and light-emitting diode (LED) placements. 
Next, the applications of the VLP systems are overviewed. Finally, this paper outlines open issues, challenges, and opportunities for the research field. 
In a nutshell, this paper constitutes the first holistic survey on VLP from state-of-the-art studies to practical uses.
\end{abstract}

\vspace{-0cm}

\vspace{-0cm}
\section{Introduction}
\label{sec:intro}
\IEEEPARstart{L}{ocation}-aware service is one of the most important functions of existing wireless networks, which enables a series of key applications such as the Internet of Things (IoT), automatic drive, and Industry 4.0 \cite{Saad2020vision6G,8808168,9815195}.
While wireless communication has seen remarkable advancements in recent decades, attaining low-cost and accurate indoor positioning through wireless systems remains a significant challenge.
With the advent of the intelligent era, the demand for 
location-aware services intensifies. 
Various indoor positioning systems like radio-frequency (RF), ultra-wideband (UWB), WiFi, Zigbee, and others have emerged.
While the existing indoor positioning technologies can either achieve accurate or low-cost positioning, it is still challenging to balance the cost and positioning performance.
For instance, WiFi is one of the most widely deployed low-cost indoor wireless technology. However, WiFi can only achieve 1 to 7 meter positioning accuracy \cite{Zhu2018AOA}.
In contrast, UWB can achieve accurate positioning at a relatively high cost \cite{song2011survey}.
%

In this context, visible light positioning (VLP) with visible light communication (VLC) has gained increasing attention. VLP utilizes ubiquitous, economical light-emitting diodes (LEDs) as transmitters, which consistently flick at a speed that cannot be detected by humans and transmit unique positioning identification (ID) information.
The popularity of LEDs makes VLP an easy-to-deploy choice in extensive applications since almost all indoor scenarios need to deploy LEDs, and many of them, such as offices, supermarkets, and subways, require LEDs for day illumination.
Furthermore, owing to the directional propagation of visible light, its limited coverage necessitates a significant number of LEDs to ensure adequate illumination. However, this multitude of LEDs contributes to providing ample positioning information within indoor spaces. 
Moreover, due to the propagation characteristics of visible light, the signals are barely interfering with each other, and thus the receiver has more accurate measurements such as received signal strength (RSS) for positioning.
Therefore, VLP can typically achieve accurate positioning due to the deterministic correspondence between the position and the received signals.

At the receiver, a commercial off-the-shelf camera or a photodiode (PD) can be used to capture the visible light signals.
These devices—PDs and cameras—are widely integrated into smart terminals such as smartphones, tablets, and vehicles. These extensively equipped hardware devices can be used to receive visible light signals from LEDs by proper software drivers without additional hardware cost.
By receiving the signals from one or multiple LEDs, the terminals employ positioning algorithm to estimate their precise location, underscoring the cost-effectiveness of VLP.
In addition, VLP stands out as a secure option for electromagnetic-sensitive scenarios such as airplane cabins and hospitals.
The combination of its cost-efficiency, utilization of prevalent hardware, and electromagnetic safety makes VLP a promising solution to tackle the ``last meter" positioning challenge.
%

\subsection{Historical View of VLC and VLP}
\subsubsection{Development of VLC}
While VLC has garnered significant attention in recent years, the utilization of light for communication traces back centuries. 
From ancient times, light has served as a medium for transmitting information, evident in early practices like torch signals and smoke signals.
Since the earliest times, the light was already used in information transmission, such as torches and smoke signals.

In the modern era, the roots of VLC extend back to the 1880s in Washington, D.C., with Alexander Graham Bell's invention of the photophone.
This groundbreaking device modulated sunlight to transmit speech, and the transmission distance can  reach to several hundred meters.
Fast forward over a century later, the advent of LED technology sparked a new era for VLC, presenting a fresh frontier for technological advancement.

VLC based on LED dates back to 1999, when researchers from Hong Kong University first proposed to modulate visible light signals on LED \cite{pang1999led,  pang1999visible}, and the proposed system has been used for intelligent transportation.
Subsequently, in 2000, the Nakagawa Laboratory in Keio University, Japan, in collaboration with Sony's Computer Science Laboratories, introduced the concept of employing visible light LEDs for indoor data transmission \cite{tanaka2000wireless}. 
They successfully established a data transmission system using a white LED bulb within an indoor environment \cite{komine2004fundamental}, marking a pivotal moment in VLC research in the 21st century.

Following these developments, the first standardization for VLC, IEEE 802.15.7, was spearheaded by the IEEE 802.15 working group in 2011. 
The same year witnessed Harald Haas demonstrating Li-Fi in a TED Talk, and Li-Fi was named by Time magazine as one of the top 50 inventions.
This milestone propelled VLC into an extensive realm of study encompassing modulation technology, multiple input multiple output (MIMO) technology, networking, and beyond \cite{matheus2019visible}.
Simultaneously, the exploration of VLC applications garnered increasing attention, extending into diverse fields including indoor access, positioning technology, and even underwater communication. The subsequent section will delve deeper into the history of VLP, outlining its evolution and early accomplishments.

\subsubsection{Development of VLP}
As VLC emerged, it spurred the development of VLP, a highly promising indoor positioning technology. This section will delve into the evolution of VLP, highlighting key milestones and achievements during its initial phases of development.

The origins of VLP can be traced back to as early as 2001, when Pang and Liu modulation of location codes within an LED location beacon system \cite{pang2001led}. This innovative system used a complementary metal-oxide semiconductor (CMOS) vision sensor to extract location code for calibration of a vehicle positioning system, which may consist of a Global Positioning System (GPS), inertial navigation system (INS), and other sensors. 
While serving as an auxiliary positioning scheme, this work marked the pioneering introduction of LED beacons into positioning methodologies.
The first dedicated VLP system tailored for indoor positioning emerged through the work of Horikawa \textit{et al.} from Japan \cite{Horikawa2004Pervasive}.
This system employed optical intensity modulation for LED illumination, transmitting visible light signals received by an image sensor. To compute the receiver's position, at least three LEDs and two image sensors were required, assuming the receiver was oriented toward the ceiling.
Then, in 2008, Yoshino \textit{et al.} \cite{yoshino2008high} proposed to use a ``collinearity condition" to calculate not only the position but also the orientation of the receiver. This condition stipulated that the projection of each spatial point and the center of the lens of the receiver must be on the same straight line.

Following the inception of VLP, researchers began to investigate innovative methods to enhance the performance of VLP.
The earliest positioning techniques within VLP \cite{pang2001led, Horikawa2004Pervasive,yoshino2008high,rahman2011high} predominantly centered around image sensing, often coupled with trilateration or triangulation methodologies.
For instance, the authors used two image sensors to establish the geometrical relationship between the distance and the position difference of the LED images and to calculate the distances between the center point of two lenses and three LEDs \cite{rahman2011high}. 
In a preliminary study \cite{rahman2011high}, researchers employed two image sensors to establish a geometric relationship among the positional differences observed in LED images. Subsequently, distances were computed by analyzing the interplay between the center points of the lenses and three LEDs.
The three distances were used to formulate the equations for location estimation according to triangulation. 
Subsequently, researchers identified and capitalized on geometric attributes such as circular features \cite{zhang2017single, zhu2023doublecircle} and rectangular features \cite{bai2021vp4l}, leveraging these distinctive shapes to further refine the accuracy and efficiency of VLC-based positioning techniques.

In 2010, a pioneering step was taken with the introduction of the initial fingerprinting algorithm for VLP \cite{hann2010white}. This innovative algorithm introduced the correlation sum ratio (CSR), a novel value derived from RSS information obtained from four LEDs. CSR functioned as distinct fingerprints corresponding to various locations. Utilizing pre-assigned information obtained offline, a receiver could determine its position by analyzing CSRs received at any given location.
Subsequently, Vongkulbhisal \textit{\textit{et al.}} \cite{vongkulbhisal2012finger} formally categorized this approach as the fingerprinting-based positioning method in VLP. Their work expanded on this concept, processing multiple signals from six transmitters to generate the fingerprint map in the offline stage or estimate its location in the online stage.

Then, the technique based on RSS was introduced for VLP in 2011 \cite{cossu2011visible}; however, it can only achieve a one-dimensional (1D) or two-dimensional (2D) positioning. Another RSS-based VLP scheme \cite{zhou2012indoor} was proposed in 2012. The proposed scheme used four LEDs and assumed Lambertian model to formulate a transmission equation group that can be solved for three-dimensional (3D) location estimation. Since then, the RSS method has been extensively used in VLP, which later gave rise to another method known as received signal strength ratio (RSSR).

Meanwhile, researchers developed a time difference of arrival (TDOA) approach for VLP in 2011. For instance, Bai \textit{\textit{et al.}} \cite{bai2011visible} proposed a TDOA-based VLP technique that can determine a vehicle's position by using photodiodes LED traffic lights. 
This approach leveraged TDOA measurements between the traffic light signals and two photodiodes, formulating equations to estimate position. They proposed two distinct methods tailored for scenarios involving one traffic light and those with two traffic lights.
In addition, Jung \textit{\textit{et al.}} \cite{jung2011tdoa} also developed a TDOA-based indoor positioning algorithm. This method synchronized the frequency address of each LED. 
Using the Hilbert transform, they extracted the phase difference to estimate distances, subsequently employing these distance values to determine the receiver's position.

Furthermore, the pioneering VLP technique utilizing angle of arrival (AOA) was introduced in 2012 \cite{lee2012location}. This method presented an AOA estimation algorithm tailored for pinpointing the location of LEDs within a VLC environment. Leveraging a circular PD array, the system aimed to accurately determine AOA, supplemented by a truncated weighting algorithm designed to bolster AOA estimation precision.
In 2014, Yamazato and Haruyama used image sensors to detect AOAs of light emitted from visible light transmitters, which were further used in pose estimation \cite{yamazato2014image}. The researchers also systematically introduced the pose and position estimation method by introducing the computer vision method.

As the field of VLP advanced, efforts intensified toward integrating two or more techniques in VLP to enhance the accuracy and practicability of indoor positioning \cite{yamazato2014image, yang2014AOA_RSS, bai2019camera}.
Notably, these endeavors included amalgamating AOA with image sensing \cite{yamazato2014image}, blending RSS with AOA \cite{yang2014AOA_RSS}, and integrating RSSR with image sensing \cite{bai2019camera} for more comprehensive pose and position estimation methodologies.

\begin{figure*}[t]
\centering
\includegraphics[height = 16cm,trim=10 20 10 8,clip]{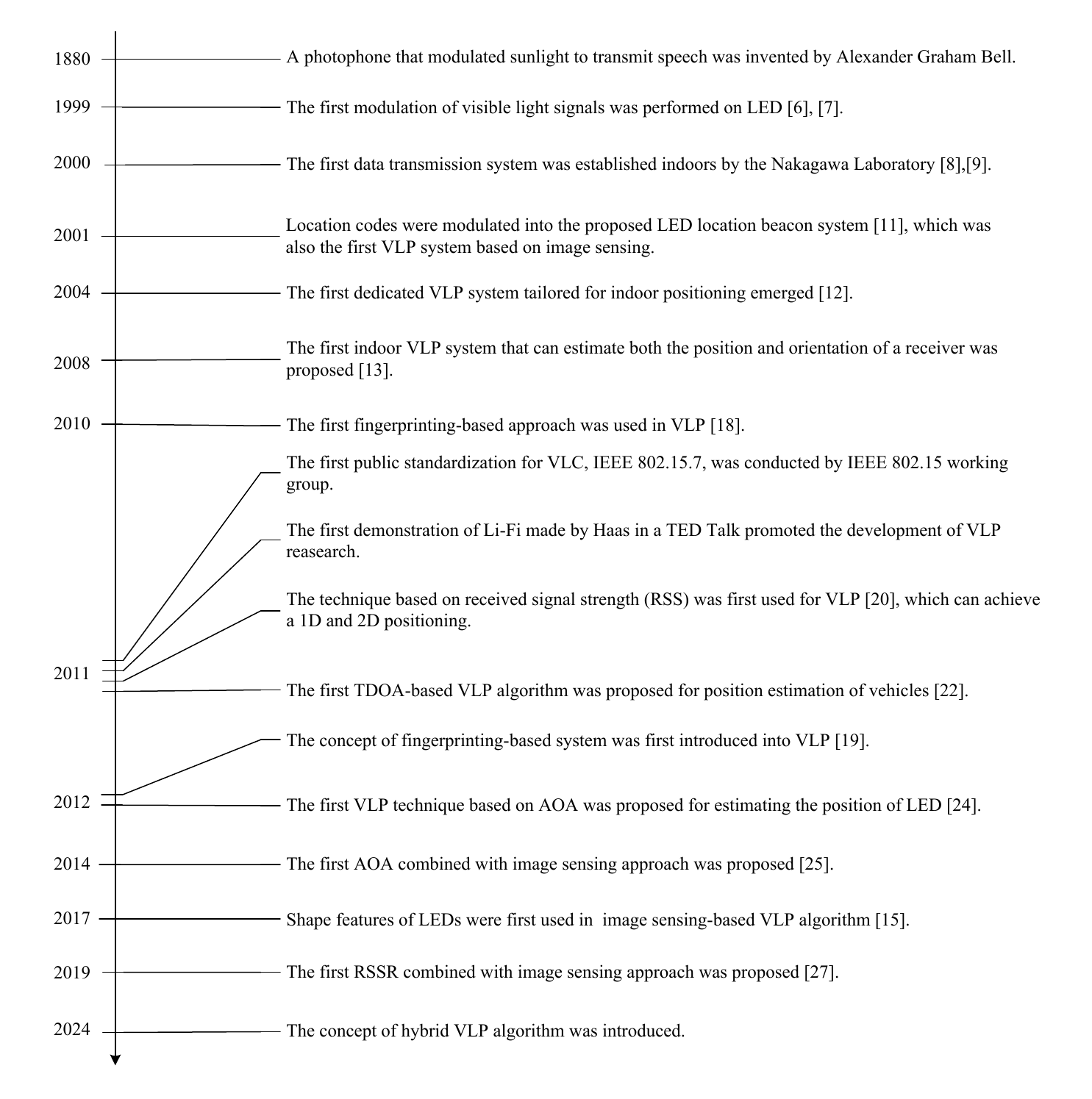}
\caption{A historical view.}\label{fig:timeaxis}
\end{figure*}

To offer a clear overview, we have encapsulated the historical evolution of VLC and VLP in Fig. \ref{fig:timeaxis}. 
Despite having about two decades of development, VLP has already achieved centimeter-level high positioning accuracy and is one of the most promising positioning technologies in indoor positioning. 
In Section \ref{VLPtech}, we will delve into the principles and latest advancements across diverse VLP techniques to provide a comprehensive understanding.

\subsection{Existing VLP Survey Papers}

Previous surveys and tutorials on VLP \cite{Zhang2013Comparison, Hassan2015Indoor, Do2016An, Luo2017Indoor, zhuang2018asurvey, jiao2017visible, Maheepala2020light} have contributed significantly to this field. For instance, Zhang and Kavehrad \cite{Zhang2013Comparison} offered a broad overview of VLP algorithm principles and the factors influencing their performance. However, this survey did not extensively delve into specific studies within each algorithm type.
Hassan \textit{et al.}  \cite{Hassan2015Indoor} classified the existing VLP systems into PD-based systems and camera-based systems, and
comprehensively investigated VLP algorithms in both systems. Do and Yoo \cite{Do2016An} provided an in-depth survey on VLC-based positioning
systems ranging from pioneering works to the state-of-the-art works, as well as the multiplexing techniques, current issues, and the research trends.
Luo \textit{et al.} \cite{Luo2017Indoor} introduced LED technology and modulation schemes and provided an updated literature review of the VLP systems.
Zhuang \textit{et al.} \cite{zhuang2018asurvey} provided a clear distinction between PD-based VLP systems and camera-based VLP systems, and the parameters
such as field of view (FoV) and Lambertian orders were also explicitly discussed.


\begin{table*}[htbp]
  \centering
  \caption{Comparison of Previous Works with This Survey.}
\resizebox{\textwidth}{!}{
    \begin{tabular}{llllllllll}
    \hline
    \textbf{Features} & \textbf{\cite{Zhang2013Comparison}} & \textbf{\cite{Hassan2015Indoor}} & \textbf{\cite{Do2016An}} & \textbf{\cite{Luo2017Indoor}} & \textbf{\cite{zhuang2018asurvey}} & \textbf{\cite{jiao2017visible}} &\textbf{\cite{Maheepala2020light}} & \textbf{This Survey} \bigstrut\\
    \hline
    \hline
    \textbf{VLC fundamentals} & No    & No & General & Explicit & Explicit & General & No& General \bigstrut\\
    \hline
    \textbf{Interaction between VLC and VLP} & No   & No    & No    & No    & No    & No  & No  & Explicit \bigstrut\\
    \hline
    \textbf{Classical VLP algorithms} & General & Explicit & Explicit & Explicit & Explicit & General  & Explicit  & Explicit \bigstrut\\
    \hline
    \textbf{Hybrid VLP algorithms} & No     & No    & Explicit & No    & General & No & No & Explicit \bigstrut\\
    \hline
    \textbf{VLP-based Heterogeneous Positioning Systems} & No     & No    & No    & Mentioned    & General & Mentioned & No & Explicit \bigstrut\\
    \hline
    \textbf{VLP network design} & No    & General & General     & No    & No    & Mentioned & General  & Explicit \bigstrut\\
    \hline
    \textbf{Performance comparisons by simulations} & No   & No & No & No & No & No & No & Yes \bigstrut\\
    \hline
    \textbf{Coverage Analysis} & No    & No    & No    & No    & No    & No  & No  & Yes \bigstrut\\
    \hline
    \textbf{Application discussion} & No    & General & No   & No    & General & No  &No  & Explicit \bigstrut\\
    \hline
    \end{tabular}%
  \label{tab:comparison}%
}
\end{table*}%

While extensive literature surveys have explored VLP algorithms and systems, they exhibit certain limitations. Current surveys primarily offer a general overview of principles, modulation schemes, and LED technologies in VLC. However, given that VLP is inherently tied to VLC, the integration between the two necessitates deeper investigation.
Moreover, emerging hybrid algorithms in VLP remain unaddressed in these surveys. Despite the significant advantages VLP offers, complexities in some scenarios prompt the need for multiple positioning technologies. 
Regrettably, existing surveys overlook this critical need for comprehensive coverage. In complex scenarios, where precise positioning becomes challenging, employing a single technology may prove insufficient. This necessitates the integration of multiple positioning technologies, an area inadequately explored by current surveys. Practical issues, such as network design and LED placement, crucially impact the availability and performance of VLP systems but have been overlooked by existing surveys that predominantly focus on multiplexing techniques \cite{Hassan2015Indoor, Do2016An, Maheepala2020light, zhuang2018asurvey, Luo2017Indoor}. 
Network designs, especially LED placement, warrant thorough investigation as VLP algorithms heavily rely on the number of LEDs detectable by the receiver. Additionally, introductory literature on VLP often emphasizes accuracy, cost, and complexity of existing systems, but ignoring coverage, which is equally important. To bridge these gaps, a comparison between this survey and existing surveys is summarized in Table \ref{tab:comparison}. 

\subsection{Contributions}
This paper's primary contribution lies in its comprehensive survey of VLP. Our overarching objective is to encapsulate the emerging research advancements in VLP systems, addressing the significant opportunities and challenges in practical VLP systems. To the authors' knowledge, this is the first survey that holistically gathers state-of-the-art and burgeoning research contributions spanning the foundational communication principles, intricate positioning algorithms, network design, and real-world applications of VLP. Our key contributions encompass the following. 
\begin{itemize}
\item We first overview PD- and camera-based VLC, which lay the foundation of PD- and camera-based VLP, respectively.
In particular, the principles of both PD- and camera-based VLC are first introduced. Then, we analyze the constraints brought forward by VLC on VLP.
\item We provide a detailed overview of VLP algorithms, including basic algorithms and new, advanced hybrid algorithms. 
We overview the existing algorithms from a new perspective of homogeneous VLP systems and heterogeneous positioning systems. Homogeneous VLP systems solely consist of VLP systems, while heterogeneous positioning systems integrate VLP with other positioning systems for positioning. 
For each category, we provide an introduction on their basic principles and a detailed survey of related research. Comparisons of the algorithms are also given in terms of the adopted receiver, the positioning accuracy, the coverage, and the orientation limitations.
\item We overview a broad range of VLP's applications such as 
industries, shopping malls, and
museums. In addition, the network designs that are also significant for VLP's large-scale applications are summarized, including multiple-access technologies, resource allocation, and LED placement.
Then, we expose the challenges and opportunities brought forward by the use of VLP. We conclude by shedding light on the potential future works within each specific area.
\end{itemize}

The rest of this survey is organized as follows. In Section II, we introduce the basis of VLP. 
Section III presents the key types of VLP algorithms in homogeneous VLP systems and compares their performance in various aspects. 
Section IV presents the positioning algorithms in heterogeneous positioning systems. 
In Section V, we discuss the VLP network designs and highlight the factors that can significantly affect the performance of VLP, while in Section VI, we overview the applications of VLP.
In Section VII, we discuss the challenges and opportunities of VLP. Finally, we draw some key conclusions in Section VIII.
\vspace{-0cm}
\section{System Model}
VLC is a promising technology that provides high-speed and secure communication due to its abundant license-free spectrum, non-electromagnetic interference, and environmental protection \cite{6GVLC,6GVLCadvanc}. This section provides a concise overview of the VLC system model. The model is differentiated into two categories based on the type of receiver device employed: PD-based VLC and camera-based VLC. 

\vspace{-0.cm}
\subsection{PD based VLC}
In VLC, data is transmitted through the modulation of light waves from the visible spectrum, spanning wavelengths from 380 nm to 750 nm.
Typically, LEDs are used as transmitters in VLC systems, including single-color LEDs and multi-color LEDs.
The multi-color LED is packaged with multiple single-color LEDs. The most commonly used multi-color LED is red green blue (RGB) LED \cite{RGBWDM}, which can be deemed as a special multi-channel transmitter that can be used to deploy multi-carrier modulation techniques \cite{RGBLED}.
Besides, VLC systems are typically equipped with PDs or cameras as receivers. PDs can typically support high-speed communications due to their high sensitivity to the light variations. In contrast, the achievable data rate of VLC based on camera typically has a low data rate due to the limited frame rate of the camera \cite{CameraRec}. However, camera is still suitable for a series of low data rate applications such as positioning, and identification. In particular, cameras are extensively equipped in smart terminals such as smartphones and vehicles. In addition, cameras can distinguish the visible light signals from interference \cite{CameraChos}.

\begin{figure}[t]
\setlength{\abovecaptionskip}{-0cm}
\centering
\includegraphics[width=0.4\textwidth]{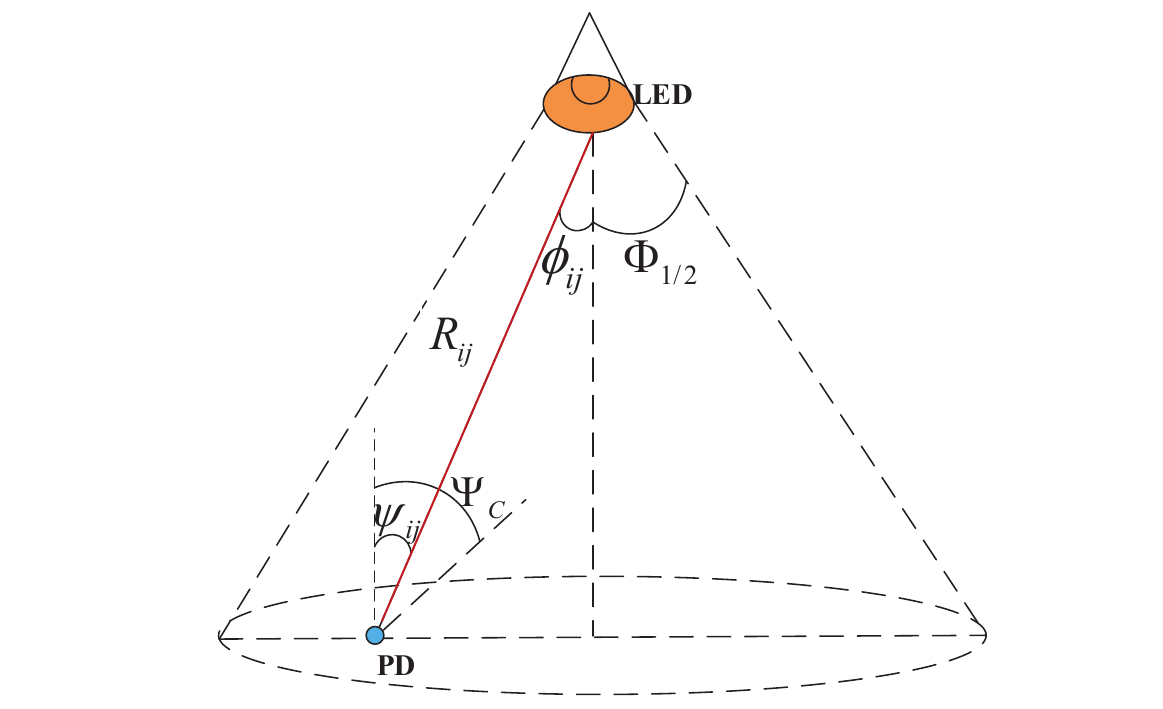}
\caption{A typical LOS channel model of VLC system.}
\label{fig:channel}
\vspace{-0cm}
\end{figure}
In VLC systems, the channel gain can be divided into the line of sight (LOS) part and the non-LOS (NLOS) part.
A directed LOS link is illustrated in Fig. \ref{fig:channel}, which can be calculated as \cite{yang2021generalized}
\begin{eqnarray}
\label{cch}
{h} = \left\{ {\begin{array}{*{20}{l}}
{\frac{{A(m + 1)}}{{2\pi {d^2}}}g(\psi ){{\cos }^m}(\phi )\cos (\psi )},&0 < \psi  \le {\Psi _{\rm{C}}}\\
{0},&{\psi  > {\Psi _{\rm{C}}}},
\end{array}} \right.
\end{eqnarray}
where $m =  - \frac{{\ln 2}}{{\ln (\cos {\Phi _{1/2}})}}$ is the order of the Lambertian emission, ${\Phi _{1/2}}$ is the semi-angle of LEDs at the half illumination power value, $A$ is the detector area, $d$ is the distance between the LED and the PD, and $g(\psi )$ denotes the gain of optical concentrator
\begin{eqnarray}
g(\psi ) = \left\{ \begin{array}{l}
\frac{{{n_r}^2}}{{{{\sin }^2}{\Psi _{\rm{C}}}}},\hspace{5pt}0 < \psi  \le  {\Psi _{\rm{C}}}\\
0,\hspace{28pt}\psi  > {\Psi _{\rm{C}}},
\end{array} \right.
\end{eqnarray}
where ${{n_r}}$ is the refractive index, ${\Psi _{\rm{C}}}$ is the receiver field of vision semi-angle.
In (\ref{cch}), $\phi $ and $\psi$ is the angle of emergence with respect to the transmitter axis and the angle of incidence with respect to the receiver axis, respectively.
Besides, the NLOS link consists of multiple NLOS paths, and the signals experience multiple reflections with a specific reflection order in each NLOS path \cite{NLOSchanl}. The reflection order refers to the number of reflections that a multipath component goes through before it reaches the receiver. Most of the existing studies use paths with order less than three to represent the complete channel to improve computation efficiency \cite{NlosLink}.

The VLC-based indoor positioning system is one of the typical applications of VLC \cite{Hassan2015Indoor, zhuang2018asurvey}.
\begin{figure}[t]
\setlength{\abovecaptionskip}{-0cm}
\centering
\includegraphics[width=0.35\textwidth]{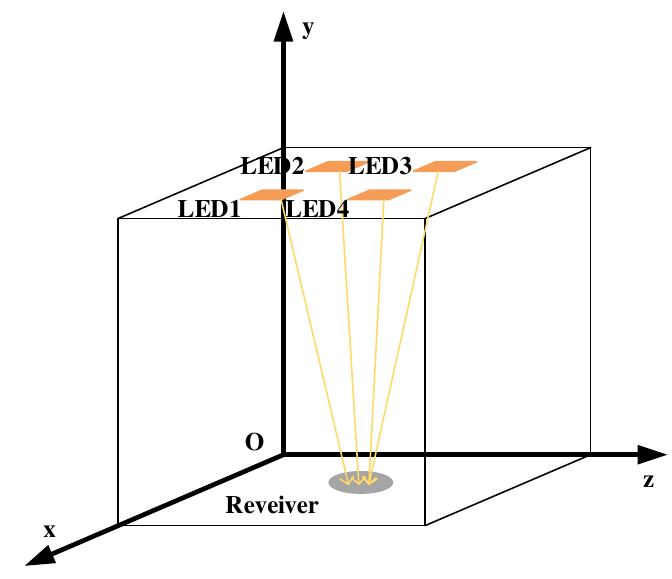}
\caption{A typical indoor multi-LED system model.}
\label{fig:system}
\vspace{-0.cm}
\end{figure}
Fig. \ref{fig:system} presents a typical indoor room equipped with multiple LEDs for lighting. Each LED has a unique location coordinates $\left( {x,y,z} \right)$ with respect to the origin ${\mathbf{O}}$, while the receiver determines its local position in the room based on the LED location coordinates. VLP is jointly achieved by the modulation module at the LED side and the demodulation modulate at the receiver side. In detail, a processor is used to encode addresses and identification (ID) numbers into bits at each LED, and then LED broadcasts its location information to the receiver. 
At the receiver, the optical signals are captured and converted into electrical current by PDs. Then, a positioning algorithm based on the measured signal is used to determine the location coordinates of the receiver. The PD-based VLP system requires a multiplexing protocol to distinguish signals emitted from different LEDs \cite{TDMaFDM}. In contrast, in the camera-based VLP system, signals emitted from different LEDs are distinguished by taking successive images at high frequencies.
In the next subsection, we will detail camera-based VLP. 

\subsection{Camera based VLC}
Camera-based VLC is equipped with cameras as receivers. The camera captures images or video streams of the intensity-modulated light sources and obtains the information by image processing \cite{le2017surveyOCC}.  

\begin{figure}
    \centering
    \includegraphics{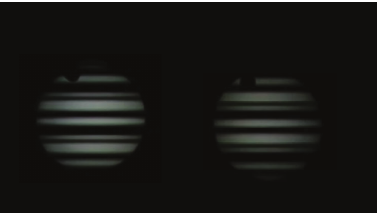}
    \caption{Fringe image captured by rolling shutter CMOS camera.}
    \label{fig:RSfringe}
\end{figure}

In the camera-based VLP systems, a CMOS camera is typically utilized, which is widely employed in mobile devices such as smartphones. The rolling shutter exposure mode of CMOS camera is used to scan a horizontal row of pixels. 
In particular, when the LED is on, the bright pixels are captured at the scanned row of pixels, whereas when the LED is off, the dark fringe is captured at the scanned row of pixels. Hence, the intensity-modulated LEDs can be captured as interleaving bright and dark fringes in a single image, as shown in Fig. \ref{fig:RSfringe}. Note that the modulation frequency of LEDs should be higher than the frame rate of the CMOS camera.  
Then, when capturing the encoded and modulated LEDs' signals as fringe images, image processing techniques are applied to assist in decoding the information. The bright and dark rows are detected by calculating pixel values. Thus, an intensity sample stream can be obtained, which will be decoded to obtain the VLC information according to the coding rules. 

Similar to PD-based VLP, the camera-based VLP also requires the LED to broadcast its location and ID information. The receiver uses the camera to receive the optical signals and locates itself by positioning algorithms. The main difference is that camera-based VLP does not rely on the channel model. It analyzes the relative position relation between the LEDs and the receiver by using imaging theory and geometric theory to achieve positioning. 

Both PD- and camera-based VLP systems typically require a direct line-of-sight VLC link between the LED and the receiver. Hence, any occlusion can lead to loss of the visible light signal. This makes VLP challenging to adapt to environments with obstacles or moving objects. In addition, PD and cameras have limited FoV, which restricts them to receive signals only in a limited area. This limitation can directly limit the coverage of the positioning. Between them, cameras usually have a wider field of view than PDs. 
Except for these commonalities, PD-based VLC is susceptible to interference between LEDs, leading to inaccuracies in positioning. Meanwhile, camera-based VLC uses the camera to capture images, and thus significant processing power is required to extract information, which may increase positioning latency.

\vspace{-0cm}
\begin{figure*}
    \centering
    \includegraphics[height=9cm]{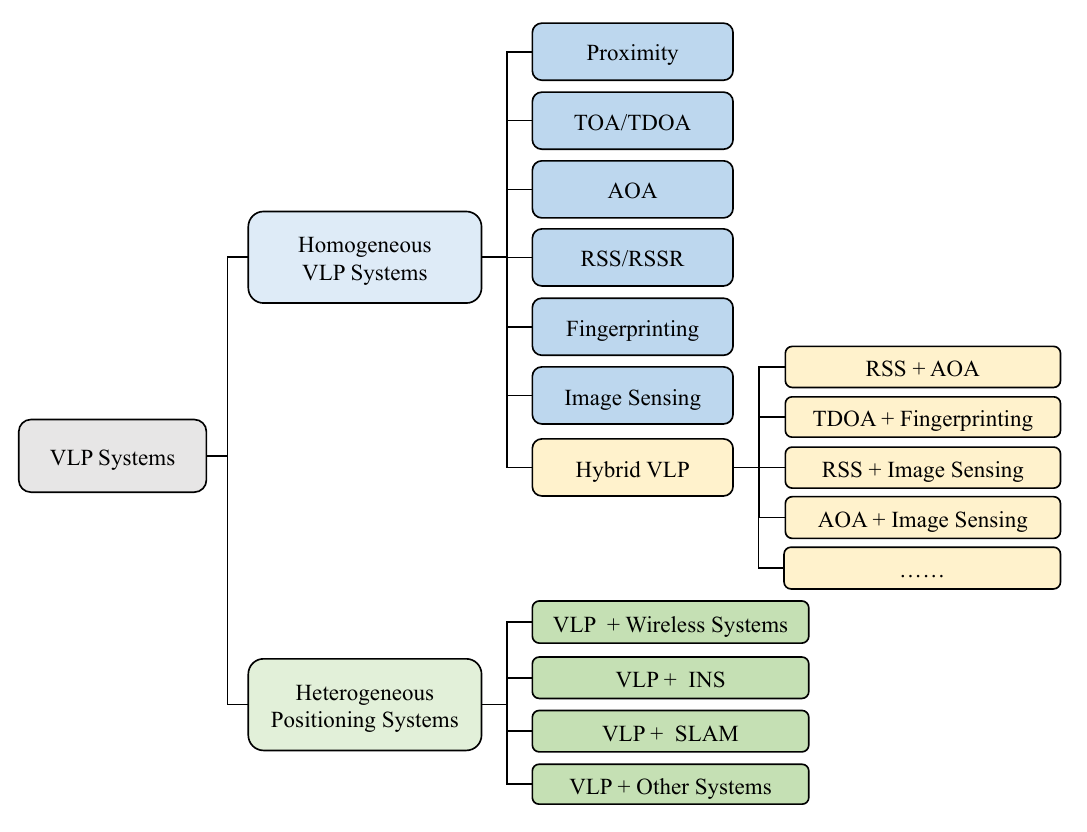}
    \caption{Taxonomy of VLP systems.}
    \label{fig:Taxonomy}
\end{figure*}

\section{Homogeneous VLP systems}\label{seq:homoVLP}

Homogeneous VLP system denotes a positioning system comprised solely of VLP systems based on VLC technologies. In this section, we first introduce the integration of VLC and VLP. Then, the positioning algorithms in the homogeneous VLP systems are detailed.

\subsection{Integration of VLC and VLP}
VLC and VLP have been identified as promising candidates to provide high-speed-data transmission and high-accuracy positioning in indoor environments \cite{vlcpTFDM,SCMvlcp,OFDMAvlcp,lin2017experimental,AIvlcp,yang2019qos,yang2020coordinated,IoTvlcp3},
where almost 80\% of mobile data traffic is generated indoors \cite{indoordata}. Most of the existing studies only focused on VLC or VLP independently, while in practical indoor environments such as offices, hospitals, supermarkets, etc., it is desirable to provide both communication and positioning services at the same time \cite{yang2019qos}. 
So far, the research on the integration of VLC and VLP (VLCP) can be divided into three categories: interference cancellation, resource allocation, and network structure design.

In many indoor environments, a large number of LEDs are used for illumination, which may lead to inter-cell interference (ICI) in VLP and VLC.
To solve this problem, the transmit time division multiplex (TDM) or frequency division multiplex (FDM) signals encoded with the unique position information of LED was appied \cite{vlcpTFDM}. A quasi-gapless integrated VLC and VLP system was experimentally demonstrated based on filter bank multicarrier-based subcarrier multiplexing \cite{SCMvlcp}. A system that realizes VLC and VLP simultaneously in the same band using orthogonal frequency division multiple access (OFDMA) was proposed \cite{OFDMAvlcp,lin2017experimental}, and the experiment results \cite{OFDMAvlcp} showed that the OFDMA base VLC positioning system provides indoor positioning, data communications and flexible bandwidth allocation.

Given that the VLP signal operates within the limited time or wavelength resources designated for VLC, several researchers investigated the resource allocation in multi-user integrated VLC and VLP systems under different quality-of-service (QoS) and positioning accuracy requirements \cite{AIvlcp}. 
Yang \textit{\textit{et al.}} \cite{yang2019qos} proposed a scheme that jointly optimized access point selection, bandwidth allocation, adaptive modulation, and power allocation to satisfy different QoS requirements.
Then, they also investigated a coordinated resource allocation approach to maximize the sum rate while guaranteeing the minimum data rates and positioning accuracy requirements \cite{yang2020coordinated}.


Moreover, in a bid to accommodate the massive connectivity needs and diverse service requirements of IoT devices, there's a proposition for a multi-layer network architecture. This architecture integrates VLC and VLP into the fifth generation (5G) networks, aiming to support massive connectivity \cite{IoTvlcp3}.

\subsection{Visible Light Positioning Algorithms}\label{VLPtech}
Figure \ref{fig:Taxonomy} demonstrates the taxonomy of the VLP systems discussed in this paper. In the homogeneous VLP systems, VLP algorithms can be classified into several categories: i) proximity, ii) time of arrival (TOA)/TDOA,  iii) AOA, iv) RSS/RSSR, v) fingerprinting, vi) image sensing, and vii) hybrid algorithm. In particular, hybrid algorithms refer to the ones that combine two or multiple VLP algorithms. Then, this section will provide a comprehensive survey for each type of the above algorithms.
%

\subsubsection{Proximity}
Proximity is the simplest positioning technique among the existing VLP techniques.  Proximity takes the location of the closest transmitter as the location of the receiver, and thus this technique can only provide an approximate position result.
In particular, each transmitter constantly broadcasts a unique ID code \cite{del2013Proximity}, which is associated with a specific location of the transmitter stored in a database. The receiver captures and detects the ID information that will be matched to the locations of the transmitters so that the receiver can estimate its location. 
When simultaneously receiving ID signals from multiple transmitters, the PD-based receiver will determine its position by the transmitter with the strongest RSS value since RSS is closely related to the distance between the transmitter and the receiver. 
To deal with the data transmission and identification in such multiple signals scenarios, 
Cherntanomwong \textit{et al.} \cite{cherntanomwong2015proximity} used TDM in their proximity-based VLP system for identification in the light overlapping area.
In addition, the camera can also be leveraged for proximity. For instance, Xie \textit{et al.} \cite{xie2018Proximity} identified the IDs of the transmitters by processing the proposed LED-ID recognition method on the captured image based on the Fisher discriminant analysis method.

Overall, proximity is simple and highly feasible in practice. However, it can only provide limited positioning accuracy, which significantly depends on the density of the LEDs.

\subsubsection{TOA/TDOA}
TOA is the absolute arrival time of signals from the transmitter to the receiver, which is a common technique for GPS \cite{leick2015gps}. In particular, TOA algorithms estimate the distances between transmitters and receivers according to the arrival time of signals, and further use the estimated distances to derive the location of the receivers. In VLC-based TOA algorithms, once the propagation delay of the signal is measured, the distance can be obtained by multiplying the delay and the speed of light. Then, TOA typically utilizes the trilateration method to obtain the position of the receiver.

\begin{figure}
  \centering
  \includegraphics[height = 4.5cm, trim = 0 10 0 0]{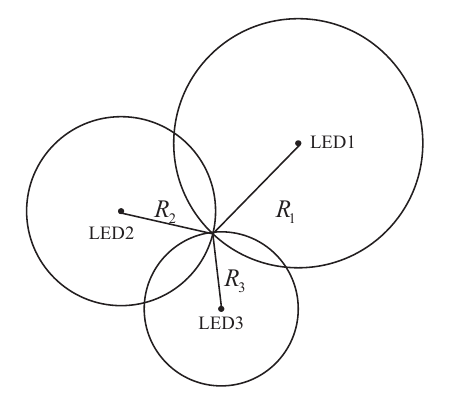}
  \caption{Positioning based on TOA.}\label{TOAfig}
\vspace{-0.4cm}
\end{figure}

Fig. \ref{TOA} illustrates the principle of the trilateration method. The LEDs locate at the center of the circles, and the radius denotes the distance between the transmitter and the receiver.  Suppose that $\left(x,y,z\right)$ is the coordinate of the receiver, $\left(x_i,y_i,z_i\right), i=1,2,...,n$ are the coordinates of $n$ LEDs, and $\tau_i, i=1,2,...,n$ are the times of arrival. Then, the distance from LED $i$ to the receiver can be expressed as
\begin{equation}\label{TOA}
 {{d}_{i}}=\sqrt{{{\left( x-{{x}_{i}} \right)}^{2}}+{{\left( y-{{y}_{i}} \right)}^{2}}+{{(z-{{z}_{i}})}^{2}}}=c\cdot {{\tau }_{i}}
\end{equation}
where $c$ represents the speed of the light prorogation. On the 2D plane, each TOA can determine a circle, and thus the location of the receiver can be determined by calculating the intersection of the three circles. In the 3D space, each TOA can determine a sphere, and thus four LEDs are needed to form four spheres and intersect at a unique point, which is the location of the receiver. Wang \textit{\textit{et al.}}  \cite{wang2013TOA} proposed a TOA positioning system, in which, the LEDs and the receiver were assumed to be synchronized perfectly. The LEDs used the orthogonal frequency division multiplexing (OFDM) technique to transmit the signals, so that the receiver can separate the signals from different LEDs and estimate the location of the receiver.

However, due to the imperfect hardware, the measured distance usually deviates slightly from the truth, and thus the drawn circles or spheres may not exactly intersect at a point. Instead, the intersection forms an overlapped area. Therefore, the least square method is applied to obtain the optimal solution \cite{liu2007survey}, and the cost function can be expressed as
\begin{equation}\label{TOAcost}
F(x,y)\!=\!\sum\limits_{i=1}^{n}{\!\left ( \!c {{\tau }_{i}}\!-\!\sqrt{{{\left( x\!-\!{{x}_{i}} \!\right)}^{2}}\!+\!{{\left( y\!-\!{{y}_{i}} \right)}^{2}}\!+\!{{(z\!-\!{{z}_{i}})}^{2}}}\right )}.
\end{equation}

\begin{figure}
  \centering
  \includegraphics[height = 5.5cm, trim = 0 10 0 0]{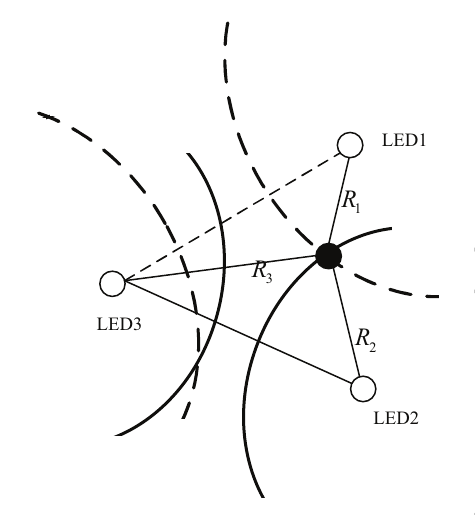}
  \caption{Positioning based on TDOA.}\label{TDOAfig}
\vspace{-0.6cm}
\end{figure}

Since the arrival time is typically rather short in VLP, TOA requires that transmitters and receivers have extremely accurate time synchronization, which is difficult to implement. 
The TDOA algorithms can alleviate the rigorous time synchronization at the receiver. They typically need at least three receivers to measure the propagation time difference between the mobile device and transmitters, and it only requires time synchronization among transmitters\cite{do2014tdoa}. Based on the time difference, the distance difference can also be derived. Letting the distances between the receiver and LED $i$ be $d_i$, and that between the receiver and LED $j$ be $d_j$, the distance difference between $d_i$ and $d_j$ can be expressed as
\begin{equation}\label{TDOA}
\begin{split}
  {{d}_{ij}}&={{d}_{i}}-{{d}_{j}} \\
 & \text{     =}\sqrt{{{\left( x-{{x}_{i}} \right)}^{2}}+{{\left( y-{{y}_{i}} \right)}^{2}}+{{(z-{{z}_{i}})}^{2}}}- \\
 & \text{     }\sqrt{{{\left( x-{{x}_{j}} \right)}^{2}}+{{\left( y-{{y}_{j}} \right)}^{2}}+{{(z-{{z}_{j}})}^{2}}}=c\left( {{\tau }_{i}}-{{\tau }_{j}} \right). \\
\end{split}
\end{equation}
According to (\ref{TDOA}), a hyperbolic positioning method can be used to determine the location of the receiver \cite{nah2013tdoa, do2014tdoa}. The positioning principle of TDOA is shown in Fig. \ref{TDOAfig}. Each distance difference can determine a hyperbola, where the two LEDs are the two focal points, and three hyperbolas corresponding to three LEDs can intersect at a point, which is the 2D location of the receiver. Similarly, the 3D location can be determined by four LEDs. 

T. H. Do \textit{\textit{et al.}} \cite{do2014tdoa} proposed a positioning system based on TDOA values. In the system, the receiver was a single PD, which received pilot signals from the LEDs. The TDOA values of the pilot signals are used to estimate the location of the receiver. The proposed system can be employed easily since the receiver can achieve positioning without embedded ID information at the LEDs. 
In addition, J. H. Y. Nah \textit{\textit{et al.}} \cite{nah2013tdoa} considered additive white Gaussian noise (AWGN) in the TDOA-based VLP system, and they proposed a Fuzzy logic algorithm and a Spring model to minimize the noise affect after position estimation.
Then, a practical TDOA-based VLP system that used a virtual local oscillator to replace the real local one was proposed \cite{Du2018tdoa}, which could reduce the hardware complexity. This system also applied cubic spline interpolation to the correlation function to reduce the rigorous requirement on the sampling rate and enhance the time resolution of cross correlation. 
Pergoloni \textit{\textit{et al.}} \cite{Pergoloni2017metameric} proposed wavelength-based localization and color-based localization mechanisms by combining traditional RSS and TDOA approaches. They assumed that each anchor point used a unique spectral signature on the wavelength domain so that the receiver could identify it and compute its location through RSS or TDOA approaches.

After that, the combination of TDOA with other received information, such as RSS and fingerprinting,  was applied in the VLP systems \cite{Keskin2018Direct, zhang2021tdoafinger}. 
Sheikh \emph{et al.} \cite{sheikh2021tdoa} proposed a TDOA-based positioning system using VLC links in a network simulator. This system computed the cross correlation between signals arriving at the target node from all sensors to formulate the distance equations between the transmitter and the target. Under the assumption that the beacon nodes are stationary and well synchronized, the proposed TDoA-based VLC scheme outperformed
the existing WiFi schemes. 
Cao \emph{et al.} \cite{cao2023visible} proposed a phase-difference-based TDOA method, in which, a convolution neural network-based phase difference estimation was designed to relieve the influence of different error sources such as Gibbs phenomenon and time synchronization error. Then, a motion-based particle filter was proposed to improve the accuracy and robustness further. The simulation results showed that an average accuracy of 0.280 m was achieved.

There is a scarcity of applications of the TOA and TDOA methods in VLP. The limited adoption of TOA/TDOA in VLP can be primarily attributed to the stringent requirement for precise time synchronization. Visible light has a much shorter wavelength than radio waves, which may necessitate more precise and sophisticated hardware for accurate time difference measurements.

\subsubsection{AOA}
\begin{figure}
  \centering
  \includegraphics[height = 4cm, trim = 0 10 0 0]{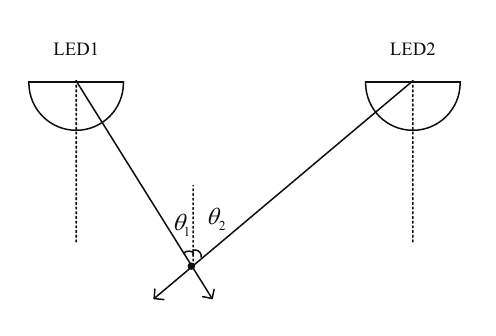}
  \caption{Positioning based on AOA.}\label{AOAfig}
\vspace{-0.6cm}
\end{figure}

The AOA algorithms estimate the location of the receiver based on the angles between the transmitters and receivers. Fig. \ref{AOAfig} illustrates the principle of AOA algorithm. When the AOA information from multiple VLC links is obtained, the location of the receiver can be determined as the intersection of VLC links according to geometric relationships. 
Suppose that the angles of incidence of the receiver from two LEDs are $\theta_1$ and $\theta_2$, respectively, and the coordinates of the receiver and the transmitters are $\left(x,y\right)$ and $\left(x_i,y_i\right)$, respectively. We have
\begin{equation}\label{AOA}
\tan {{\theta }_{i}}=\frac{y-{{y}_{i}}}{x-{{x}_{i}}},i=1,2,...
\end{equation}
Then, triangulation can be used for AOA algorithms according to (\ref{AOA}) \cite{yang2014AOA_RSS, Eroglu2015AOA, Hong2020AOA}. In particular, at least two transmitters are needed for 2D AOA positioning algorithm, while at least three transmitters are needed for 3D positioning. 
Sun \textit{\textit{et al.}} \cite{Sun2015AOA} provided a traditional AOA algorithm, and the authors derived Cramer-Rao bound to analyze the theoretical accuracy of the algorithm. Yu \textit{\textit{et al.}} \cite{Yu2021AOA} employed an optimal optical omnidirectional angle estimator for AOA system, which can mitigate the side effect of the uncertain orientation of the receiver. When more than two LEDs were exploited, two supplemented methods that selected the LEDs with more reliable measured power data were proposed to reduce the estimation error. In addition, Soner \textit{\textit{et al.}} \cite{Soner2021AOA} proposed a quadrant photodiode-based AOA algorithm for vehicle positioning to avoid collision and platooning.

The performance of AOA algorithms significantly depends on the acquisition of AOA values. AOA values are often calculated by the VLC signals received by PDs and the image captured by the camera. 

The first approach estimates AOA based on the Lambertian channel model, which is relative to radiation angle and incidence angle according to (\ref{cch}). Therefore, it is typically assumed that the transmitter plane should be horizontal and parallel to the receiver plane \cite{Eroglu2015AOA, Sun2015AOA, Yu2021AOA}, which is inapplicable to scenarios with random receiver orientation. To overcome this limitation, a positioning framework was proposed based on angle differences of arrival in a 3D coordinates system \cite{Zhu2018AOA}, which had no receiver orientation limitations. Some researchers investigated different designs of AOA detectors \cite{zhu2019CPDs, Stefanie2017visible, zhang2022beacon} to relax the orientation limitation at the receiver. For instance, Zhu \textit{\textit{et al.}} \cite{zhu2019CPDs} proposed to use pairs of PDs, namely complementary PDs, to construct AOA estimators. Stefanie \textit{\textit{et al.}} \cite{Stefanie2017visible} investigated a new form of AOA detector that was a quadrant PD placed below a transparent aperture in an opaque screen. Zhang \textit{\textit{et al.}} \cite{zhang2022beacon} used two optical AOA estimators for locating LEDs. The AOA estimators had fixed relative positions, and each estimator consisted of four PDs with different orientations.
In addition, MEMS sensors like accelerators and gyroscopes were also used to measure the orientation of the receiver, including azimuth, roll, and pitch angles \cite{ Hong2020AOA}. 

There are also approaches that use the camera to estimate the position of the receiver. They obtain the three angles by calculating the trigonometric relationship between the coordinates of the transmitters and the receivers from the captured image. 
For instance, Hossei \textit{\textit{et al.}} \cite{Hossei2019GOPA} proposed a geometrical optics positioning algorithm (GPOA) based on AOA to locate smartphones. 

\subsubsection{RSS/RSSR}
The RSS algorithm is one of the most widely-used methods in indoor VLP, and it estimates the location of the receiver based on the power of the received signal, i.e., RSS values from LEDs. RSS values can be easily obtained using PDs equipped on the receiver. 
As illustrated in (\ref{cch}), the RSS increases as the distance between the transmitter and the receiver decreases. When the angles of incidence and radiation are fixed, the distance between the receiver and the corresponding LED can be calculated using the detected RSS values based on the Lambertian channel model. Then, the trilateration method can be applied for positioning.

Gu \textit{\textit{et al.}} \cite{gu2014RSS} proposed to calculate the horizontal coordinates of the receiver given the RSS information from four LEDs, after which the height of the receiver was also estimated. It also used Kalman and particle filters to realize target tracking. An RSS-based trilateration method using the code division multiple access technique was proposed by Guan \textit{\textit{et al.}} \cite{guan2017RSS}. Assuming that the transmitter and the receiver planes were parallel, they used RSS to estimate the distance by increasing the height of the transmitter from 0, and the distance was used to calculate the location of the receiver using the trilateration method. 
Zhang \textit{et al.} \cite{zhang2019high} proposed a deep neural network (DNN)-based RSS positiong system. The DNN was trained by the Bayesian regularization based on the Levenberg-Marquardt algorithm, so that unknown positions across the same area can be estimated by using the trained DNN on limited training points. 
Then, Saboundji \textit{\textit{et al.}} \cite{saboundji2022accurate} implemented the artificial neural network learning for an RSS algorithm to achieve highly accurate and efficient indoor positioning. Salman \textit{\textit{et al.}} \cite{salman2022performance} suggested the demonstration of a 3D VLP system, combining RSS and trilateration solution so that the speed of the computation may be increased considerably. Shen \textit{\textit{et al.}} \cite{shen2022hybrid} used hybrid maximum likelihood/maximum a posteriori principle for a multiple LEDs - multiple PDs system to achieve positioning, which took into account the presence of prior information on the orientation. 
%

In addition, RSSR algorithms are also applied to VLP. Different from RSS algorithms that directly use the received power to calculate the distance between the transmitter and the receiver, RSSR algorithms calculate the ratio of distances according to the ratio of the received power from multiple transmitters and estimate the location based on the distance ratios. RSSR has the advantage that the ratio of received power can avoid the error caused by the non-zero irradiance \cite{Do2016An}, and thus it has less error than the distance directly obtained from the power. However, RSSR algorithms typically require the receiver plane to be parallel to the transmitter plane \cite{jung2013RSSR,jung2014RSSR}, so that the incidence angle can be equal to the irradiance angle. Hence, the received power can be derived from (\ref{cch}) as
\begin{equation}\label{Pr_RSSR}
{{P}_{r}}={{P}_{t}}\frac{(m+1)A}{2\pi {{d}^{2}}}g(\psi){{\cos }^{m}}(\phi ){\cos }(\psi )=\frac{U}{{{d}^{m+3}}}
\end{equation}
where $U=\frac{(m+1){P_t}A g(\psi)h^{m+1}}{2\pi }$ is a constant that can be calculated. Then, the distance ratio can be expressed as
\begin{equation}\label{d_ratio_RSSR}
\frac{{{d}_{1}}}{{{d}_{2}}}=\sqrt[{}^{{}^{m + 3}}] {\frac{{{P_{r_2}}}}{{{P_{r1}}}}} .
\end{equation}
By substituting the coordinates of two LEDs, $({{x}_{1}},{{y}_{1}},{{z}_{1}})$ and $({{x}_{2}},{{y}_{2}},{{z}_{2}})$ into (\ref{d_ratio_RSSR}), we have
\begin{equation}\label{RSSRfunction}
\frac{\sqrt{{{\left( x-{{x}_{1}} \right)}^{2}}+{{\left( y-{{y}_{1}} \right)}^{2}}+{{(z-{{z}_{1}})}^{2}}}}{\sqrt{{{\left( x-{{x}_{2}} \right)}^{2}}+{{\left( y-{{y}_{2}} \right)}^{2}}+{{(z-{{z}_{2}})}^{2}}}}=\frac{{{d}_{1}}}{{{d}_{2}}}.
\end{equation}
In this way, two simultaneous equations can achieve 2D positioning when three LEDs are detected. Three simultaneous equations with four LEDs used can achieve 3D positioning.

Since RSS or RSSR algorithms rely on a perfect channel model, they are susceptible to the incidence and irradiance angles. Several RSS and RSSR algorithms have been proposed to tackle the transmitter and receiver orientation limitations \cite{wang2017RSSR_LED, wang2017RSSR_PD, li2019positioning}. 
Wang \textit{\textit{et al.}} proposed two designs for RSSR-based VLP algorithms, including multiple directional LED array \cite{wang2017RSSR_LED} and multiple direction PD array \cite{wang2017RSSR_PD}, to reduce the error caused by the orientation of transmitters and receivers, respectively. 
In addition, Li \textit{et al.} \cite{li2019positioning} also considered the orientation uncertainty of the receiver by coping with the non-linear relationship between the RSS and the orientation uncertainty. They utilized the first and second-order Taylor series expansion of RSS to find an accurate approximation for the RSS when the orientation of the receiver was uncertain.

\subsubsection{Fingerprinting}
In fingerprinting algorithms, one or multiple features related to the receiver's position are selected as fingerprints. The position of the receiver is estimated by matching the measured data with the prestored location-related data. In particular, there are two phases for the fingerprinting algorithm, including the offline phase and the online phase. In the offline phase, the data related to the location are collected and stored in the database. In the online phase, the location of the target is estimated by matching the currently measured data to the pre-stored database.

There are various features of the signal that can be selected as fingerprints. Most of the existing literature adopted RSS as the fingerprints, and they typically used RSS vector to further enhance the positioning accuracy\cite{Bakar2020finger, Alam2018finger, Abou2021finger, Guo2017finger}. RSS vector consists of several RSS values from multiple transmitters \cite{Arfaoui2021finger, Zhao2017finger,Chen2018finger}.
Received signal strength indicator (RSSI), as another representation of RSS, was used as a fingerprint \cite{Wei2017finger, Cui2020finger}. In addition, the channel impulse responses (CIR) were also selected as a fingerprint \cite{Hossei2020finger}.
Yang \textit{\textit{et al.}} \cite{yang2013finger} chose the extinction ratio that represented the ratio of received powers when bit 1 and bit 0 are transmitted as the fingerprint.
In another work \cite{vatansever2017finger}, light power distribution calculated from grayscale images was used as fingerprints for indoor positioning and tracking. 
Moreover, Zhao \textit{\textit{et al.}} \cite{zhao2017navilight} proposed a LightPrint as the fingerprint, which is a vector of multiple light intensity values obtained from existing lighting infrastructure with any unmodified light source during the user's walks.

To match the measured data to the database accurately, a number of match methods have been studied. In particular, probabilistic methods and k-nearest neighbors (kNN) algorithms are extensively used.

The probabilistic method utilizes the probability distribution of RSS to estimate the location of the target with the Bayesian method. Supposing that the transmitters are independent of each other, the probability of a location candidate $L_i$ can be calculated as the probability of the RSS from all transmitters by
\begin{equation}\label{Bayes1}
P\left( \boldsymbol{R}|{{L}_{i}} \right)=\prod\limits_{j=1}^{n}{P\left( {{R}_{j}}|{{L}_{i}} \right)}
\end{equation}
where $\boldsymbol{R}=\left( {{R}_{1}},{{R}_{2}},...,{{R}_{n}} \right)$ is an RSS vector composed of RSS values from $n$ transmitters, and  $P\left( {{R}_{j}}|{{L}_{i}} \right)$ represents the probability of the RSS value $R_j$ from the $j$th transmitter when the receiver locates at location $L_i$. The probability distribution of the RSS is calculated and stored in the offline stage. Then, the positioning coordinate can be estimated by a weighted average of the coordinates as
\begin{equation}\label{Bayes2}
\left( \widehat{x},\widehat{y} \right)=\sum\limits_{i=1}^{n}{P\left( {{L}_{i}}|\boldsymbol{R} \right)}\left( {{x}_{{{L}_{i}}}},{{y}_{{{L}_{i}}}} \right)
\end{equation}
in the online stage. 
Kail \textit{\textit{et al.}} \cite{kail2014finger} proposed a probabilistic positioning system that addressed the problem of unpredictable obstructions and synchronization error based on a Bayesian model. Another research \cite{qiu2016finger} proposed a light-signal decomposition method to extract fingerprints, and then a Bayesian localization framework was applied to improve the precision. Ou \textit{\textit{et al.}}  \cite{ou2022gaussian} used kernel functions to model the spatial correlation nature for the reflected lights. Based on that, target location estimates are given by the Bayesian inference both in a discrete setup and a continuous framework, which are the probabilistic fingerprinting and the maximum likelihood estimate.

The kNN method is another typical method adopted in fingerprinting algorithms based on Euclidean distance \cite{Bakar2020finger, Alam2018finger, Abou2021finger, Guo2017finger, Arfaoui2021finger}. The Euclidean distance $D_j$ between the online measured RSS vector $\boldsymbol{R}=\left( {{R}_{1}},{{R}_{2}},...,{{R}_{n}} \right)$ and the offline RSS vector ${{\boldsymbol{S}}_{j}}=\left( {{S}_{j,1}},{{S}_{j,2}},...,{{S}_{j,n}} \right)$ can be calculated by
\begin{equation}\label{knn}
{{D}_{j}}=\sqrt{\sum\limits_{i=1}^{n}{{{\left( {{R}_{i}}-{{S}_{j,i}} \right)}^{2}}}}
\end{equation}
where ${S}_{j,i}$ denotes the prestored RSS value from the $i$th transmitter in RSS vector, ${{\boldsymbol{S}}_{j}}$, which is related to the $j$th location candidate. According to the Euler distance, there are usually several location candidates can be selected. When $k$ location candidates are selected, the final location of the target is determined by averaging $k$ location candidates. For simplicity, $k=1$ was typically selected \cite{vongkulbhisal2012finger, Hossei2020finger}.

Although the kNN method is simple and highly feasible, it is limited by the size of the cell, i.e., the density of the grid. To address this issue, weighted K-nearest neighbors (WkNN) was adopted for fingerprinting algorithms \cite{Bakar2020finger, Alam2018finger, Abou2021finger, Cui2020finger, oh2022vlc}. The WkNN method assigns weights to the distances so that the neighbor with a smaller distance has a greater weight compared with the neighbor with a greater distance. Based on the WkNN method, Cui \textit{et al.} \cite{Cui2020finger} proposed a clustering algorithm and used Spearman distance to improve the performance of the fingerprint algorithm. 
Oh \textit{et al.} \cite{oh2022vlc} constructed a fingerprinting database based on the channel characteristics of VLC and obtained the approximate location of the user by applying WkNN. In addition to kNN, extreme learning machine (ELM) and random forest were also applied to fingerprinting \cite{Guo2017finger}.

\subsubsection{Image Sensing}
With the popularity of the camera equipment, the image sensing-based VLP technique has also attracted massive attention. This technique estimates the location of the target by analyzing the geometric relation between the LEDs' coordinates and their corresponding projection on the image. The image sensing in VLP is similar to the multi-view geometry in computer vision. The key difference is that the camera can receive the VLC signals to obtain the information of transmitters in VLP.

To identify the transmitters and obtain the information in image sensing algorithms, the VLC signals should be received and decoded. Thus the under-sampled phase shift on-off keying \cite{zhang2019image} based modulation and camera rolling shutter effect \cite{fang2017image, guan2019image} based modulation methods are adopted for communications.

\begin{figure}
  \centering
  \includegraphics[height = 5.5cm, trim = 0 10 0 0]{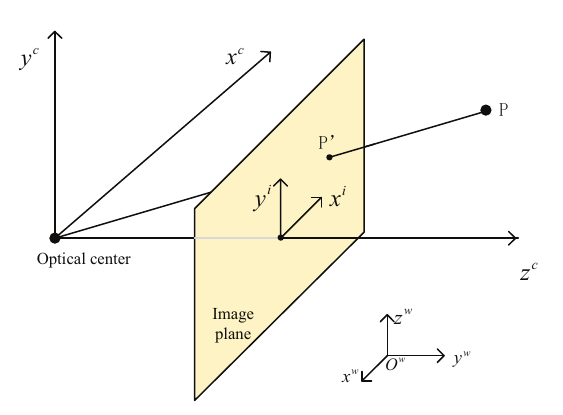}
  \caption{Pinhole camera model.}\label{imagefig}
\vspace{-0.6cm}
\end{figure}

In image sensing, there are four coordinate systems established for calculating the geometric relation between the corresponding points. They are a 3D world coordinate system, a 3D camera coordinate system, a 2D image coordinate system, and a 2D pixel coordinate system. The image coordinate $\left(x^{\rm{i}},y^{\rm{i}}\right)^{\rm{T}}$ can be easily obtained through image processing \cite{li2018vlc}, while the world coordinate of the LED $\left(x^{\rm{w}}, y^{\rm{w}}, z^{\rm{w}}\right)^{\rm{T}}$ can be obtained by receiving the modulated VLC signal from the LED. The camera coordinate $\left(x^{\rm{c}},y^{\rm{c}},z^{\rm{c}}\right)^{\rm{T}}$ can be calculated by 
\begin{equation}\label{image1}
{{\left( {{x}^{\text{i}}},{{y}^{\text{i}}}, 1 \right)}^{\text{T}}}=\mathbf{A}{{\left( {{x}^{\text{c}}},{{y}^{\text{c}}},{{z}^{\text{c}}}, 1 \right)}^{\text{T}}}
\end{equation}
according to pinhole camera model as illustrated in Fig. \ref{imagefig}, where
\begin{equation}
    \mathbf{A }= \begin{bmatrix}
   f & 0 & 0 & 0  \\
   0 & f & 0 & 0  \\
   0 & 0 & 1 & 0  \\
\end{bmatrix}
\end{equation}
is the calibration matrix of the camera, which can be calibrated and known in advance, and $f$ is the focal length of the lens. Then the geometric relation between the world coordinate system and the image coordinate system can be described as
\begin{equation}\label{image2}
{{\left( {{x}^{\text{i}}},{{y}^{\text{i}}}, 1 \right)}^{\text{T}}}=\mathbf{A} \cdot \left( \begin{matrix}
   \mathbf{R} & \mathbf{T}  \\
   \mathbf{0} & \mathbf{1}  \\
\end{matrix} \right){{\left( {{x}^{\text{w}}},{{y}^{\text{w}}},{{z}^{\text{w}}}, 1 \right)}^{\text{T}}}
\end{equation}
where $\mathbf{R}$ is a  $3\times 3$ rotation matrix representing the pose of the receiver, and $\mathbf{T}$ is a $3\times 1$ translation vector representing the location of the receiver. The goal of image sensing algorithms is to find the pose and location of the receiver, i.e., the camera.

Single view geometry is widely used in image sensing based VLP algorithms \cite{guan2019image, li2018vlc, kim2016image, lain2018image, zhang2017single, kuo2014luxapose, yang2015wearables}. Guan \textit{\textit{et al.}} \cite{guan2019image} proposed a double-LED positioning (DLP) system using CMOS image sensor. The authors utilized a rolling shutter mechanism and machine learning algorithm to identify the IDs of LEDs. DLP required the camera to parallel to the LEDs and solved the symmetry problem in this circumstance. Lain \textit{\textit{et al.}} \cite{lain2018image} proposed a $K$-pairwise LED image-sensor-based VLP (IS-VLP) algorithm for indoor positioning.
$K$-parwise LED IS-VLP identified the mapping between the received LED IDs and their images to determine the coordinates of the LEDs, which was then used to estimate the location of the receiver.
Moreover, Huang \textit{\textit{et al.}} \cite{huang2022three} proposed a 3D NLOS VLP system using a single LED and an image sensor (IS) to address the problem of obstructed LOS paths. Two virtual LEDs reflected from the ground together with the real LED are captured for positioning. 

In addition, the researchers also gradually focus on the geometric features captured by the image sensing. For instance, Zhang \textit{et al.} \cite{zhang2017single} extracted the feature of a single circular LED to locate the target by assuming a weak projection model with the assistance of another built-in sensor. Zhu \textit{\textit{et al.}} \cite{zhu2023doublecircle} proposed a VLC-assisted perspective circle and arc algorithm (V-PCA), which exploited the geometric features between two LEDs and their circle images for positioning. To reduce the required luminaires to one, they further proposed a visual odometry-assisted VLP algorithm \cite{zhu2024visible}. 
Meanwhile, Bai \textit{\textit{et al.}} \cite{bai2021vp4l} considered the rectangular features of a single luminaire in the positioning process.
%

%

\subsubsection{Hybrid VLP algorithms}

To attain practical and highly accurate positioning, recent research has focused on merging multiple VLP algorithms, termed hybrid VLP algorithms in this context. 

These approaches often amalgamate two classical VLP algorithms to achieve superior positioning performance. We categorize these hybrid algorithms into several distinct types for clarity and classification.

\paragraph{RSS and AOA hybrid algorithm}
The combination of RSS and AOA has been studied \cite{yang2014AOA_RSS, othman2021RSSAOA, sahin2015RSSAOA, hou2015RSSAOA, aparicio2022experimental}, since both RSS and AOA values can be simply detected by PDs. Yang \textit{\textit{et al.}} \cite{yang2014AOA_RSS}, used a multiple-PD structure to obtain RSS and AOA values simultaneously. Then 3D positioning was achieved based on RSS and AOA values without the limitation of the tilted angles at the receiver. Othman \textit{\textit{et al.}} \cite{othman2021RSSAOA} also utilized RSS and AOA values simultaneously and applied a weight least square estimation to find the location of the target. In addition, an RSS-based approach using a nonlinear least squares (NLS) estimator was proposed \cite{sahin2015RSSAOA}. This approach also developed an analytical learning rule based on the Newton-Raphson method to reduce the complexity of the NLS estimator, which used AOA-based localization as an initial point for the learning rule. Hou \textit{\textit{et al.}} \cite{hou2015RSSAOA} proposed to use one LED lamp to position, and a smartphone was used to receive RSS and AOA to estimate its 3D location. Aparicio \textit{\textit{et al.}} \cite{aparicio2022experimental} presented an AOA-based triangulation algorithm that used the acquired RSS values to estimate the image points for each LED and then implemented a Least Squares Estimator (LSE) and trigonometric considerations to estimate the receiver's position. 

\paragraph{TDOA and fingerprinting hybrid algorithm}
An indoor VLP method combining TDOA and fingerprints was proposed \cite{zhang2021tdoafinger}. A visible light fingerprint database was first built, and then the TDOA algorithm was used to determine the application range of the fingerprint. The final location of the target was obtained via WkNN. A TDOA-based VLP is proposed for locating maritime targets by combining with fingerprinting in harbor-border inspection \cite{feng2022time}. An unmanned aerial vehicle was used as the receiver with five PDs installed. The TDOA measurements were used to formulate a weighted least squares positioning problem based on the Chan-Taylor (CT) method. Then, after a coarse target location, a fingerprinting positioning method was proposed to improve the accuracy and robustness.

\paragraph{RSS and image sensing hybrid algorithm}
Given the prevalence of commercial mobile devices equipped with both cameras and PD, researchers have explored combining information captured by these components for positioning purposes. 
For instance, Bai \textit{et al.} \cite{bai2019camera} proposed a camera-assisted RSSR (CA-RSSR) positioning algorithm by jointly using the camera and PD in the smartphone to position. The authors used the image captured by the camera to calculate the incidence angle of the receiver, which facilitated accurate positioning without limitation of the orientation at the receiver. To improve the accuracy of this hybrid positioning system, the authors further proposed an enhanced CA-RSSR (eCA-RSSR) algorithm \cite{bai2021ecarssr} to mitigate the error caused by the distance between the PD and the camera. 
Meanwhile, they also proposed a location method \cite{bai2020received} that integrates visual and RSS information from three LEDs, irrespective of their orientations, and receivers. By analyzing images captured by the camera, they derived incidence angles through geometric principles. Combining these angles with RSS values, they estimated the desired irradiance angles and distances between LEDs and the receiver, subsequently estimating the receiver's location.
Furthermore, Hua \textit{et al.} \cite{hua2021fusionvlp} proposed a FusionVLP system based on the fusion of the measurements from a PD and a low-cost camera, in which an Ensemble Kalman Filter (EnKF) module was used for real-time positioning, and a Fixed-Lag Ensemble Kalman Smoother (FLEnKS) module was used for semi-real-time positioning. To improve the accuracy, they introduced a new two-step adjusted PD RSS model, including interference adjustment and Lambert Coefficient adjustment.

\paragraph{AOA and image sensing hybrid algorithm}
With the geometric relation between the transmitters and their projections on the image plane, AOA is also exploited in image sensing-based VLP algorithms \cite{Hossei2019GOPA, Cincotta2019, liu2022visible}, and the AOA value can be obtained by the projection model of the camera.
For instance, Hossei \textit{et al.} \cite{Hossei2019GOPA} proposed GPOA based on AOA using the front-facing camera of the smartphone. GPOA designed space-color-coded identifiers and balanced the number of different colored LEDs to keep the lighting white. Then, GPOA used the similarity relation in the projection model to estimate the location of the target. 
In addition, luminaire reference points (LRPs) and hybrid imaging-photodiode (HIP) receivers were investigated for VLP \cite{Cincotta2019}, in which, AOA was measured by the HIP receiver. The receiver can be located when only one luminaire is in the FoV of the receiver by precisely defining multiple LRPs on the single luminaire. 
Liu \textit{et al.} \cite{liu2022visible} proposed to fuse AOA and TDOA with a Time of Flight camera as the receiver. This proposed fusion algorithm leveraged a hybrid Chan/Taylor series expansion method \cite{Hao2011combination} for positioning, and the experimental results demonstrated that the fusion algorithm can achieve more accurate positioning compared to using a single algorithm.

We summarize and compare the above VLP algorithms in Table \ref{tab:VLP algorithm}. We mainly compare the number of LEDs required in 3D positioning, whether the auxiliary device is required, and the advantages and weaknesses of each algorithm. 
For instance, the existing image sensing research recognizes that image sensors can use more visual information, but it needs to consider the problem of positioning delay. 
Compared with the traditional VLP algorithms, the hybrid VLP algorithms are dedicated to solving the challenges of VLP in practice, such as the orientation limitation at the receiver and the requirement of multiple LEDs. However, this approach also increases the complexity of the receiver devices since they often need the PD and image sensor simultaneously at the receiver. 
\begin{table*}[htbp]
\newcommand{\tabincell}[2]{\begin{tabular}{@{}#1@{}}#2\end{tabular}}
  \centering
  \caption{Summary of homogeneous VLP algorithms}
  \resizebox{\linewidth}{!}{
  
  \begin{threeparttable}
  
    \begin{tabular}{lcccll}
    \hline
    {\textbf{VLP Algorithm}} & {\textbf{\tabincell{c}{Minimum of \\ required LEDs\tnote{*} }}} & {\textbf{\tabincell{c}{Multipath \\ effect}}} & {\textbf{\tabincell{c}{Auxiliary \\ devices}}} & {\textbf{Advantage}} & {\textbf{Weakness}}\\
    \hline
    \hline
    Proximity &  1 & No & No & \tabincell{@{}l}{a) Simple to implement \\ b) Low cost} & \multirow{1}{*}{Low accuracy}\bigstrut[b]\\
    \hline
    TOA/TDOA & 4 & Yes & Yes & High accuracy & \tabincell{@{}l}{a) High cost \\ b)  Rely on time Synchronization \\ c) Complicated to implement } \bigstrut[b] \\
    \hline
    
    AOA & 3 & Yes & Yes & \tabincell{@{}l}{ a) Do not rely on channel model \\ b) Do not take up bandwidth} & \tabincell{@{}l}{a) High cost \\ b) Time Synchronization \\ c)  Complexity } \bigstrut[b] \\
    \hline

    RSS & 4 & Yes & No & \tabincell{@{}l}{ a) Low cost \\ b) Simple to measure RSS} & \tabincell{@{}l}{ a) Rely on channel model \\ b) Receiver need to be parallel to LEDs} \bigstrut[b] \\
    \hline

    Fingerprinting & 1 & No & No &  \tabincell{@{}l}{ a) High accuracy \\ b) Low requirements at the receiver } & \tabincell{@{}l}{a) Need to build and update \\ fingerprint database \\ b) Poor portability  } \bigstrut[b] \\
    \hline

    Image sensing & 2 & No & No & \tabincell{@{}l}{ a) High accuracy \\ b) Multiple visual information \\ available } & \tabincell{@{}l}{a) Slower response than PD \\ b) Poor portability \\ c) Sensitive to environmental changes } \bigstrut[b] \\
    \hline

    Hybrid VLP & - & - & - & \tabincell{@{}l}{ a) Relax the orientation limitation \\ at the receiver \\ b) Fewer LEDs required } & \tabincell{@{}l}{Increase the complexity at the receiver   } \bigstrut[b] \\
    \hline


    \end{tabular}%

 \begin{tablenotes}
        \footnotesize
        \item[*] Note that the minimum of required LEDs represents the typical requirement of most algorithms of the current VLP algorithm type, which does not include the algorithms using additional sensor devices or special receiver and transmitter devices.  
      \end{tablenotes}
    \end{threeparttable}
}

  \label{tab:VLP algorithm}%
\end{table*}%

\subsection{Comparisons of VLP algorithms}
This subsection evaluates the positioning accuracy and the coverage ratio of several representative VLP algorithms. 
In particular, we consider the following four algorithms, the RSS positioning algorithm, the DLP algorithm \cite{guan2019image}, the eCA-RSSR algorithm \cite{bai2021ecarssr}, and the V-PCA algorithm \cite{zhu2023doublecircle} to represent the traditional trilateration method, image sensing, hybrid VLP algorithm, and a novel image sensing-based algorithm. The RSS-based positioning algorithm leverages PD as the receiver, while the DLP algorithm and V-PCA algorithm leverage the camera as the receiver. The eCA-RSSR algorithm uses the PD and camera simultaneously as the receiver for positioning.


V-PCA takes the visual features of the LED's circle contour into consideration and requires two LEDs for 3D positioning. The rest of the three algorithms treat the LEDs as point light sources. DLP, eCA-RSSR and RSS require two, three, and four LEDs to achieve 3D positioning, respectively. 
In addition, RSS and DLP require the receiver to be parallel to the transmitter, while V-PCA and eCA-RSSR can estimate the orientation directly. 
For fairness, the performances of the above algorithms are all simulated in a room of 5 m $\times$ 5 m $\times$ 3 m. There are four LEDs placed at the locations of (1.75 m, 1.75 m), (1.75 m, 3.25 m), (3.25 m, 3.25 m), and (1.75 m, 1.75 m) with the height of 3 m. The radius of the LEDs is 10 cm. 
The four LEDs transmit the location information of the LED's center. In eCA-RSSR, DLP, and RSS, the geometric center of the LED is treated as equivalent to the LED itself for modeling and calculation.

\begin{figure}
    \centering
    \includegraphics[height = 6.5cm]{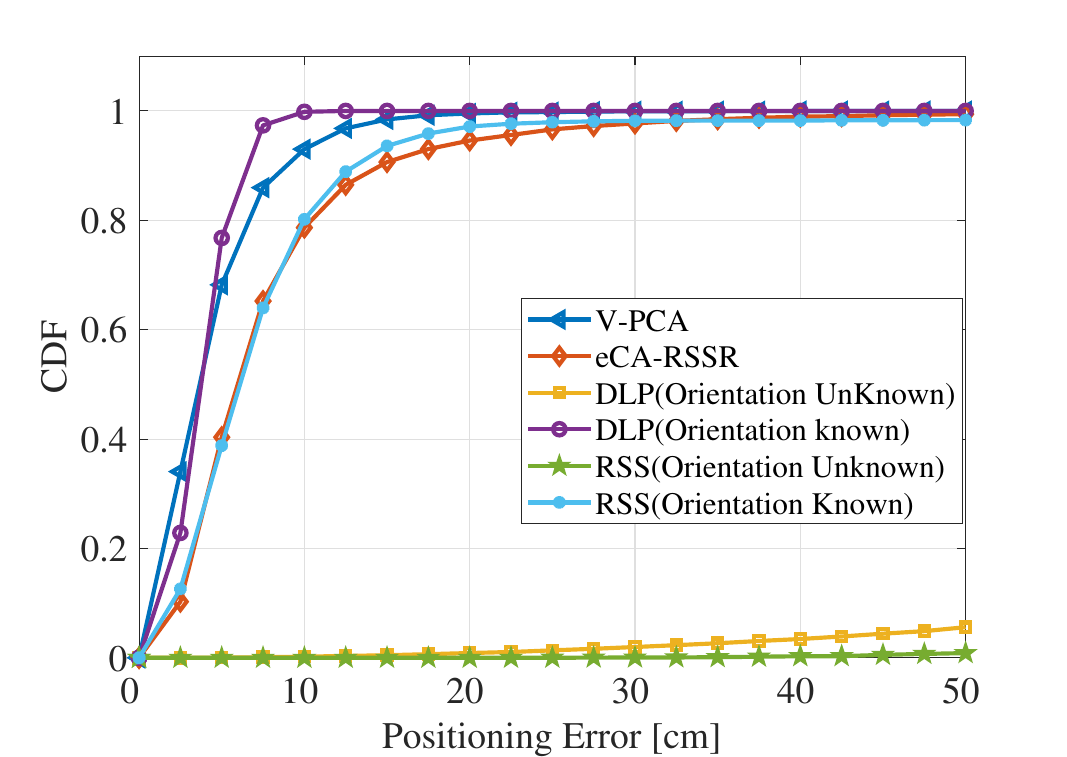}
    \caption{The CDF of position estimation error.}
    \label{accuracy}
\end{figure}

Figure \ref{accuracy} compares the positioning accuracy of the four algorithms in terms of cumulative distribution function (CDF) curves. 
There are 10,000 location samples generated randomly in the room with random tilted angles. We simulate the performance of the RSS and DLP algorithms when the orientation of the receiver is known and unknown.
From Fig. \ref{accuracy}, it can be observed that when the orientation is unknown, the V-PCA algorithm can achieve a 93\% accuracy of about 10 cm. The eCA-RSSR algorithm can achieve a 78\% accuracy of about 10 cm. However, RSS and DLP cannot work in this circumstance. 
When the orientation of the receiver is given, DLP achieves a 98\% accuracy of about 10 cm, while RSS achieves an 80\% accuracy of about 10 cm.  
In this circumstance, DLP and RSS even outperform V-PCA and eCA-RSSR, respectively. This is because an estimation error exists in the process of estimating the orientation of the receiver in V-PCA and eCA-RSSR. 

\begin{figure}
    \centering
    \includegraphics[height=6.5cm]{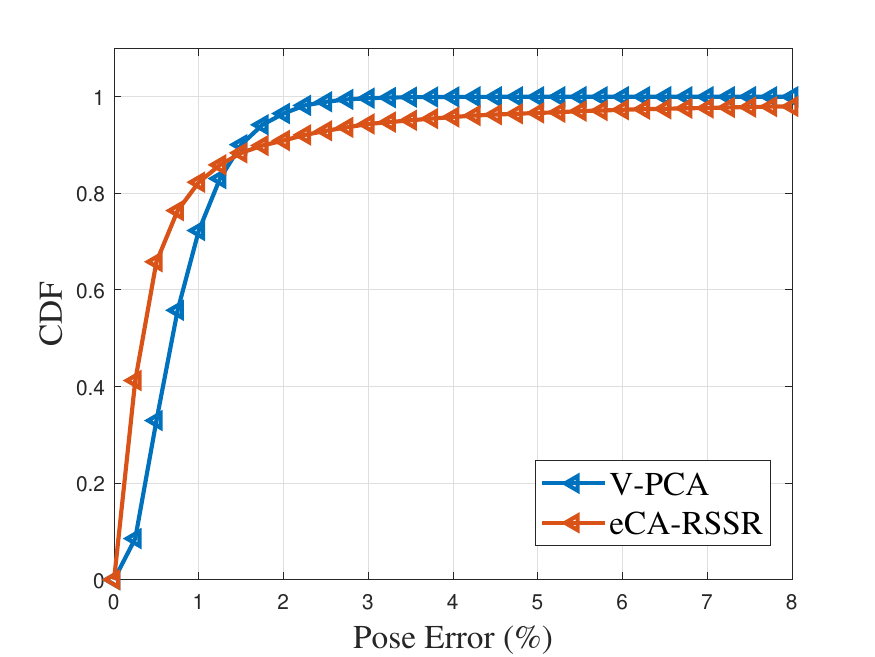}
    \caption{The CDF of pose estimation error.}
    \label{fig:rot}
\end{figure}

In addition to the location estimation error, the positioning algorithms also focus on the pose estimation error. Fig. \ref{fig:rot} compares the pose estimation accuracy of V-PCA and eCA-RSSR algorithms using CDF curves. Since RSS and DLP algorithms require the receiver to be parallel to the transmitter, they are not included in this comparison. 
We use ${E_\textrm{pos}}\left( \%  \right) = \left\| {{{\boldsymbol{q}}_{{\rm{true}}}} - {\bf{q}}_{\rm{est}}} \right\|/\left\| {\boldsymbol{q}}_{\rm{est}} \right\|$ \cite{zhu2023doublecircle} to evaluate the pose estimation accuracy, where ${{\boldsymbol{q}}_{{\rm{true}}}}$ and ${\boldsymbol{q}}_{\rm{est}}$ are the normalized quaternions of the true and the estimated rotation matrices, respectively. 
It can be observed that V-PCA exhibits similar performance to eCA-RSSR. In particular, V-PCA can achieve 2\% pose estimation error with 96\%, while eCA-RSSR can achieve 2\% pose estimation error with 91\%. 

To sum up, the V-PCA algorithm has the best performance in terms of positioning accuracy and robustness since it needs only two LEDs and has a relaxed orientation limitation at the receiver. In addition, DLP and RSS are sensitive to the tilted angles of the receiver. 
In this regard, future work could explore adaptive algorithms that dynamically adjust to the environment and receiver conditions to optimize positioning accuracy and robustness.

\begin{figure}
    \centering
    \includegraphics[height = 6.5cm]{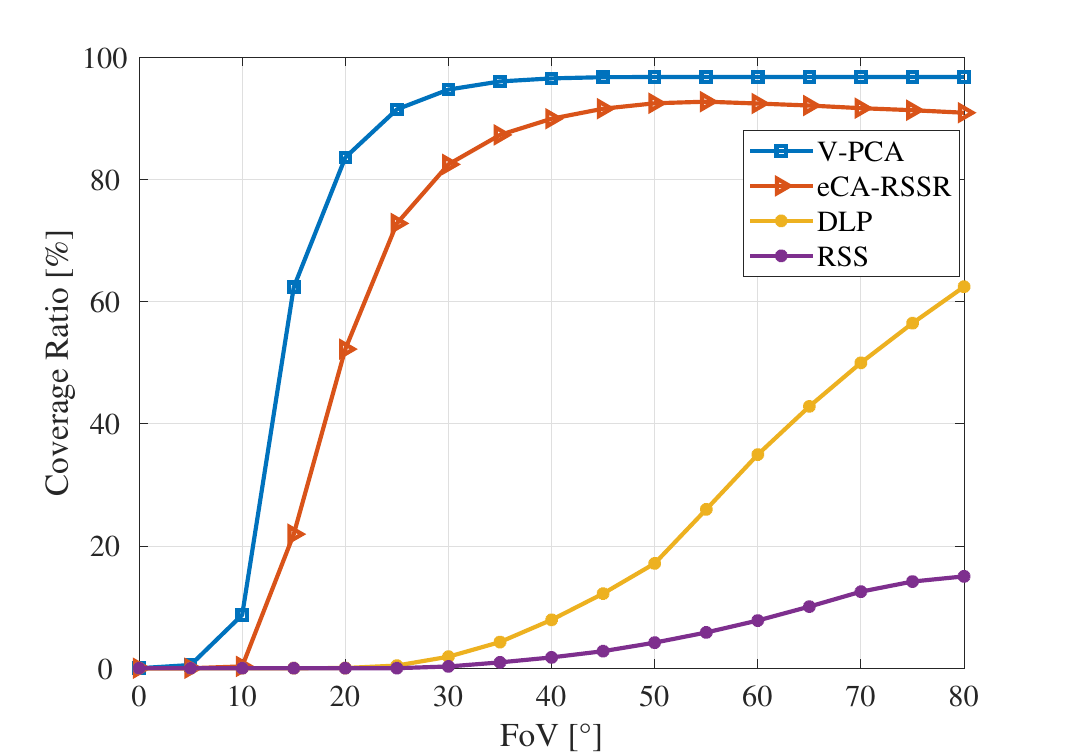}
    \caption{The coverage ratio.}
    \label{fig:cover}
\end{figure}

Figure \ref{fig:cover} shows the coverage ratio of 3D positioning by positioning algorithms versus the FoV of the receiver. The coverage ratio is calculated by \cite{bai2021ecarssr}
\begin{equation}\label{eq:cover}
    R_{\rm{cov}} = \frac{A_{\rm{effective}}}{A_{\rm{total}}} \times 100 \%,
\end{equation}
where $A_{\rm{effective}}$ represents the effective area that the receiver can locate itself, and $A_{\rm{total}}$ represents the total area of the room.  
In Fig. \ref{fig:cover}, FoV varies from  $0^\circ$ to $80^\circ$. The location samples are chosen along the length, width, and height of the room, with a 5 cm separation from each other. For RSS and eCA-RSSR, an SNR of 13.6 dB is considered to ensure reliable communication. 
In addition, RSS and DLP both require the receiver to be parallel to the transmitter, while V-PCA and eCA-RSSR do not. Hence, for V-PCA and eCA-RSSR, the receiver is tilted to test how many LEDs can be available at each location sample. For RSS and DLP, the receiver is always oriented vertically upward.
From Fig. \ref{fig:cover}, it can be seen that V-PCA always achieves the highest coverage ratio for all FoV values. It performs consistently well from $25^\circ$ to $80^\circ$, whose $R_{cov}$ exceed 90\% and about 20\% better than eCA-RSSR. 
This is because eCA-RSSR needs three LEDs for positioning, while V-PCA only needs two, although both eCA-RSSR and V-PCA have no strict requirements for orientation.
In addition, although DLP also only needs two LEDs for positioning, it has an unsatisfactory performance of coverage ratio when compared to V-PCA. This is because DLP requires the receiver to face up vertically. 
It can also be observed that RSS has the lowest coverage ratio among the four algorithms. This is because four LEDs and a fixed orientation of the receiver are required for RSS to achieve positioning. 
In conclusion, the coverage ratio of the positioning algorithms is related to the required number of LEDs and the orientation of the receiver. When the receiver's orientation is relaxed, the coverage ratio will be greatly improved. Additionally, a higher coverage ratio of the positioning algorithm is achieved as the algorithm demands fewer LEDs. 
This indicates that future research should focus on developing algorithms that are robust to receiver orientations and that can achieve positioning effectively with a minimal number of LEDs. This would not only enhance the coverage ratio of VLP systems but also reduce deployment costs and complexity. Furthermore, the integration of VLP with other indoor positioning technologies is a promising research direction to offer improved coverage and accuracy. The next subsection will detail the lessons learned.

\subsection{Summary and Lessons Learned}
Table II summarizes and compares different types of VLP algorithms mentioned in this section. We mainly compare the number of LEDs required in 3D positioning, the requirement of auxiliary devices, and the advantages and weaknesses of each algorithm. 
The solely VLP algorithms have their advantages and weaknesses. For instance, the RSS method has the advantage of simple measurement and implementation while having the weakness of relying upon the channel model assumption. These assumptions may not always hold in real-world scenarios, leading to potential inaccuracies. Fingerprinting has high accuracy and low requirements at the receiver. However, the fingerprint database makes this method with poor portability. 
Compared with a single VLP algorithm, hybrid VLP algorithms offer several advantages in practice, such as relaxing the orientation limitation at the receiver and reducing the number of LEDs required. However, the cost of the hybrid VLP algorithms is increased since they often need both PD and image sensor at the receiver. 


To sum up, homogeneous VLP systems face the following challenges. First, the orientation of the receiver is required to orient toward the ceiling to ensure accurate positioning. Second, the ambient light and multipath inflection interference is another significant issue, as light from external sources and inflection can interfere with the VLP signals, thus reducing the accuracy and reliability. Third, obstructions in the environment, such as walls, furniture, or even people, can block the LOS link between the transmitter and the receiver, leading to signal loss and positioning failure. Finally, the need for multiple LEDs to achieve comprehensive coverage and accurate positioning can also limit the application of VLP systems. The requirement for multiple light sources means that a dense network of LEDs is required, which can be both costly and technically challenging. To tackle these challenges, robust solutions and advanced algorithms are required to enhance the performance and feasibility of VLP systems in diverse environments.

In particular, we suggest some countermeasures below. First, machine learning (ML) \cite{9562559,9210812,chen2021communication} has the potential to address the above issues in VLP systems, such as orientation limitation and inflection light interference. 
Tran \emph{et al.} \cite{tran2022machine} reviewed the applications of ML in VLP systems and concluded that ML was a promising choice to improve the performance of VLP systems. An ML-based method \cite{yuan2018tilt} was proposed to address the problem of the tilted angle limitation at the receiver while maintaining high positioning accuracy. In addition, ML methods in VLP algorithms \cite{liu2022machine, alenezi2022machine} have been proposed to utilize the indoor reflection of optical propagation for accurate positioning. These studies demonstrate that incorporating ML into VLP systems can significantly enhance the robustness and accuracy, making VLP capable of overcoming inherent limitations. 
In addition, investigating unmodified LED-based VLP holds the potential for relaxing the reliance on perfect channel model assumptions. Some works \cite{zhao2017navilight, liang2021novel} have attempted to use unmodified LEDs for positioning and verified the feasibility.
Moreover, integrating VLP with other positioning technologies, such as Wi-Fi, INS, and simultaneous localization and mapping (SLAM), can also potentially tackle the challenges of VLP systems. In this integration, multiple-source information can be used to mitigate the orientation limitation, occlusions, and the need for multiple LEDs.

Next section will detail the integration of VLP and other positioning technologies.

\section{heterogeneous positioning Systems}\label{hybridtech}

Different from the homogeneous VLP systems, heterogeneous positioning systems amalgamate VLP with other positioning methods such as WiFi, INS, and SLAM, which have gradually emerged as a focal point in the academic.

Despite VLP offering high precision and cost-effectiveness, achieving satisfactory performance in intricate indoor settings remains challenging. The integration compensates for the shortcomings of VLP and makes the LED-based positioning systems benefit from multi-source information to achieve improved performance. For instance, when positioning services are interrupted due to inaccessible visible light signals, other integrated positioning methods can play a role and improve the adaptability of the positioning system. 
Moreover, when visible light signals are accessible, VLP can augment the accuracy of other positioning methods, and conversely, these methods can offer valuable multi-dimensional information to VLP. For example, an inertial measurement unit (IMU) can measure Euler angles and other pertinent data, thereby enriching the VLP system's information framework.
In essence, heterogeneous positioning systems are crafted to leverage the strengths of multiple positioning techniques.

In the existing research, there have been several fusion methods, such as environment-based signal fusion, weight-based signal fusion, spatial-temporal data fusion, and multi-sensor data fusion:
\paragraph{Environment-based signal fusion}
The environment-based signal fusion methods choose the appropriate signal source based on the characteristics of the environment. For instance, based on the signals received in the positioning environment, the heterogeneous positioning system can switch to visible light signals or other signals, such as WiFi and Bluetooth, ensuring stable positioning service.

\paragraph{Weight-based signal fusion}
The weight-based signal fusion methods assign weights to the integrated positioning algorithms. The weights are adjusted according to the reliability and quality of different positioning algorithms. 

\paragraph{Spatial-temporal data fusion}
In the spatial-temporal data fusion used in heterogeneous positioning systems, the positioning results of different positioning methods interact with each other. For instance, historical positioning data and motion models such as IMU are used to predict current positioning results, while VLP results can be used to correct current positioning results. 

\paragraph{Multi-sensor data fusion}
Finally, benefiting from the high degree of embedded sensors integration of existing intelligent terminals \cite{guo2019survey}, some researchers focus on multi-sensor data fusion to integrate VLP with other positioning algorithms \cite{yby9,cheng2020singlecircle,yby18,wen2023enhanced, alcazar2024seamless}. By collecting data from both VLP and other positioning, heterogeneous positioning algorithms can combine these data to achieve more accurate and reliable positioning. 

Note that the above fusion methods are not independent of each other, and the existing works may use multiple fusion methods simultaneously. Next, this section will detail and analyze the existing heterogeneous positioning systems according to the different positioning algorithms that VLP integrates with. 

\subsection{VLP Integrated With Wireless Systems}

The combination of VLP and other wireless systems, such as RF, Bluetooth, and acoustic, have been studied \cite{survey-wifi,zigbee,rf,vlp-5g,yby2,bluetooth,yby3,yby4}. 

\subsubsection{VLP and RF}
When combined with RF, a heterogeneous indoor positioning system using both LiFi and WiFi was envisaged to improve the accuracy of indoor positioning \cite{survey-wifi}.
Zigbee wireless network was constructed in a VLP system to transmit VLC data to reduce the position estimation error caused by nearby visible light channels \cite{zigbee}.
There was also a two-stage positioning system developed \cite{rf}. In the first stage, RF was used to detect the room where the device located. In the second stage, LiFi was employed to detect the specific position of the device. The estimation error was reported to be only 5.8 cm. 
In addition, Shi \textit{et al.} \cite{vlp-5g} proposed a 5G indoor positioning scheme based on VLC and indoor broadband communication for the museum. The system utilized unlicensed visible light of the electromagnetic spectrum to provide visitors with high-accuracy positioning on a mobile device, realizing a mean positioning error of 0.18 m. 

\subsubsection{PD and Bluetooth}
When combined with Bluetooth, Luo \textit{et al.} \cite{yby2} proposed a Bluetooth signal based spring model to hybrid VLP and Bluetooth positioning. The intensity of visible light signals was detected through the Bluetooth beacon set in advance to match the fingerprint database. Simulation results showed that the system can achieve an average positioning accuracy of 6 cm. 
Hussain \textit{et al.} \cite{bluetooth} used a VLC-based indoor mapping application to facilitate Bluetooth MAC address mapping. In this way, the advantages of VLC and Bluetooth can be combined to achieve superior positioning performance. 
Albraheem \textit{et al.} \cite{albraheem2023hybrid} utilized VLC proximity for initial location determination and Bluetooth RSS trilateration for refinement, achieving an accuracy of up to 0.03 meters. The VLC system estimated the receiver's location based on modulated information from the light source, which was demodulated at the receiver into a distinct identifier code. Subsequently, the Bluetooth system used an RSSI-based trilateration technique to further refine the receiver's location.

\subsubsection{VLP and Acoustic}
In addition, the combination of acoustic positioning and VLP has also been studied \cite{yby3} and \cite{yby4}. Akiyama \textit{et al.}  \cite{yby3} measured the propagation time of acoustic signal through an acoustic sensor equipped in the phone. With the obtained propagation time, TOA was then used in VLP to achieve precise localization. Experiments showed that the positioning accuracy can be achieved up to 100 to 200 mm. Png \textit{et al.} \cite{yby4} integrated two acoustic sensors on the Arduino hardware board. The sensors can detect the moving distance of pedestrians on the X and Y axes in a 2D plane respectively, which improved the positioning readings to centimeters and helped the system obtain higher accuracy.

\subsection{VLP Integrated With INS}
INS uses the measurements from inertial sensors to estimate position and orientation \cite{engelsman2023information}, and different sensors perform different measuring functions. The accelerometer can detect the velocity information in specific directions, while the gyroscope can obtain the angle information of moving targets. The magnetometer can monitor the strength of the magnetic field in the surrounding environment. 
Due to the need for the comprehensive processing of the above various motion information, complex sensors integrating the above components have been gradually produced, such as the IMU integrating the accelerometer, gyroscope, and magnetometer, and the six-axis angle sensor which can measure the direction angle and inclination angle. The appearance and development of motion sensors also directly promote the development of inertial navigation.
Due to the ubiquity of the camera, PD, and inertial sensors in existing intelligent devices, the combination of VLP and INS has also attracted increasing attention, which improves the performance of the positioning system through the assistance of inertial navigation \cite{yasir2014AOA,yby1,yby6,yby7,yby8,yby9,yby10,yby11,cheng2020singlecircle,yby13,yby14,yby15,yby16,yby17,yby18,yby19}.  

As mentioned above, the accelerometer is used to measure the acceleration of the moving object in specific directions. Yasir \textit{et al.} \cite{yasir2014AOA} proposed a positioning scheme fusing light intensity sensors and accelerometers. The accelerometer was used to determine the direction of the receiver, which can simplify the posture calculation process and reduce the average positioning error to 25 cm. Nakajima \textit{et al.} \cite{yby1} integrated magnetic positioning with VLP to construct the hybrid scheme, in which the pedestrian direction and the walking angle were measured through a magnetometer equipped with the phone and VLC system. The scheme helped travelers to estimate the LED node at the destination more accurately, and then determine its approximate location by using the nearest neighbor method of VLP. The experimental results confirmed that positioning accuracy can be improved through the assistance of geomagnetic value correction. Sertthin \textit{et al.} \cite{yby6} proposed a switching receiver position estimation scheme for a VLC-ID and 6-axes sensor-based positioning system. The positioning performance was improved by optimizing the estimated error distance, and the accuracy was improved by more than 30\% compared with the traditional positioning scheme only based on VLC-ID. Then, Wang \textit{et al.} \cite{yby7} sensed self-attitude to obtain the smartphone's normal vector by a built-in accelerometer and magnetic sensor, achieving sub-meter accuracy in typical office corridor areas. 
In addition, there was the combination of VLP with geomagnetic and gyroscopic sensors \cite{yby19}, in which, the position information was first obtained by VLP, and then the direction information of dead reckoning was detected through sensors. Finally, the multiple information was processed by the Kalman filter to provide essential data correction. Experimental results showed that the direction error of the proposed system was always less than 6 degrees.

In addition to the aforementioned heterogeneous systems, other mainstream existing heterogeneous positioning systems \cite{yby9, yby10, yby11, cheng2020singlecircle, yby13, yby15,yby16,yby17,yby18,alcazar2024seamless, wen2023enhanced, Wei2023high, tan2024indoor} that combine VLP with INS can be categorized into two types: VLP integrated with IMU-based positioning and VLP integrated with pedestrian dead reckoning (PDR).
In particular, IMUs are primarily concerned with the direct sensing of inertial forces and rotational movements. 
In addition, PDR, as a specific application of INS, leverages inertial sensors—primarily accelerometers and gyroscopes—to estimate the position of pedestrians.
The measurements of IMU and PDR can be incorporated with VLP data to achieve comprehensive and precise navigation. 

\subsubsection{VLP and IMU}
IMU can measure the acceleration and angular velocity of the object jointly in real-time, and thus it can be combined with VLP when visible light signals are not available. Liang \textit{et al.} \cite{yby9} proposed an EKF-based visual and inertial fusion method for visible light positioning with an IMU and a rolling shutter camera. The results showed that the method can keep the RMSE errors of positioning to 5 cm. Zhang \textit{et al.} \cite{yby10} developed an indoor positioning system based on VLP and sensor fusion techniques. The system enhanced positioning accuracy by fusing data collected by the image sensor and motion sensors of a smartphone. The experiment results showed that the proposed algorithm improves the position accuracy by 44\% when compared with the algorithm employing a single image sensor. Then, Qin \textit{et al.} \cite{yby11} fused the IMU and VLC information in a tightly coupled scheme so that the VLP system could continually provide location service under the aforementioned intermittent outage condition. Experimental results showed that the proposed method had better stability than the PnP methods. 
At the same time, fusion positioning schemes using a single circular LED were proposed \cite{cheng2020singlecircle, yby13}.
In particular, the geometric features of LED projection and plane intersection lines were used for deriving relative position relation between the LED and the receiver, and IMU was used to estimate the orientation of the receiver \cite{cheng2020singlecircle}. 

\subsubsection{VLP and PDR}
PDR can achieve positioning without GPS by analyzing the data obtained by inertial sensors. However, due to the accumulation of external noise and internal drift error, PDR is arduous to be adopted as an independent navigation or positioning scheme without proper correction \cite{yby14}. If PDR can be integrated with VLP, its accuracy can be significantly improved with the assistance of a VLP system. 

There have been studies on the combination of VLP and PDR\cite{yby15,yby16,yby17,yby18,alcazar2024seamless, wen2023enhanced, Wei2023high, tan2024indoor}. 
Li \textit{et al.} \cite{yby15} proposed to simultaneously process the data from PDR and VLP by particle filter. In particular, after obtaining positioning data of PDR in some specific positions, the system put it into a particle filter together with VLP data. Filter processed two kinds of data and obtained the final coordinates of pedestrians. The scheme solved the performance degradation caused by the multipath effect and light transmission-blocking and alleviated the problem of cumulative error of inertial navigation. The experimental results showed that positioning error could be as low as 0.14 m. After that, they have also proposed to use the Kalman filter to simultaneously process VLP and PDR data \cite{yby16}. The inertial navigation without requiring external sources improved the positioning performance of the VLP system in the region with weak visible light intensity or occlusion, while the accumulated error of inertial navigation was reduced with the assistance of VLP data. The experimental results showed that the hybrid system was better compared to the single VLP or inertial navigation.

Different from works using PD as the receiver \cite{yby15, yby16}, some works \cite{yby17, yby18} employed the image sensor. 
In particular, Huang \textit{et al.} \cite{yby17} completed cell recognition and 3D positioning by capturing the image of a single luminaire, realizing continuous and stable indoor positioning by using sparse light source beacons in actual scenes. Wang \textit{et al.} \cite{yby18} proposed to use mode average to process heading angle data for pedestrian positioning, which improved positioning accuracy and shortened reliable running time. In the positioning process, the LED with modulation code was used as the absolute position beacon, and then the LED image was captured and decoded by the camera to obtain the absolute position. The information would be used to periodically correct the position of the PDR, thereby avoiding the accumulation of PDR technology and obtaining high precision. 
Wen \emph{et al.} \cite{wen2023enhanced} designed a VLP-integrated PDR system for real-time localization using a common smartphone, achieving 98.5\% decoding accuracy for 20-bit IDs at 2.1 m height, even under sparse LED distribution and variable shutter speed. In the system, the VLP provided global information and periodic calibration to the PDR, while the PDR addresses the discontinuity problem caused by the limited FoV of the camera sensor.
Alcázar-Fernández \emph{et al.} \cite{alcazar2024seamless} presented a mobile application-based indoor positioning system for museums that combines smartphone IMU sensors with PDR and VLP AoA algorithms. The system estimates the user's position using the device's inertial sensors when no light source is detected. Simulation results demonstrated that this approach reduces the average error to 0.85 meters and provides accurate navigation.

\subsection{VLP Integrated With SLAM}
There are also the combination of VLP and SLAM \cite{yby20, yby21} has been studied. In particular, the work \cite{yby20} proposed a loosely coupled multi-sensor hybrid method based on VLP and SLAM. The multi-sensor hybrid system could achieve an average accuracy of 2 cm in milliseconds even when the number of received LEDs was insufficient, or the link was blocked. The work \cite{yby21} exploited VL-GraghSLAM to improve the accuracy of trajectory estimation, and a Kalman filter was used to fuse visible light fingerprint and inertial sensor data, to complete locating users. Comprehensive experiments showed that the performance of the system was better than the traditional WiFi-based fingerprint recognition method. 

Moreover, some researchers designed and implemented heterogeneous systems by using robotic operating systems (ROS) to collect and process multi-source data \cite{yby22, yby23}. 
Guan \textit{et al.} \cite{yby22} utilized the principle of double-lamp positioning and implementing locating through loosely coupled ROS nodes based on the loose coupling characteristics. The images of LEDs were captured by the industrial camera node and subscribed by the locator node, and then the position results were computed. Experimental results showed that the proposed system can provide indoor localization within 2 cm and only 0.35 seconds for positioning per time. Besides, a VLP system with an improved calibration program was proposed \cite{yby23}, which used mobile robots to collect data and obtain environmental maps with beacon position and identity by using the Google Cartographer SLAM algorithm \cite{hess2016real}. The performance of the new method was significantly improved, and the error was almost halved.

\subsection{VLP Integrated With Other Systems}
There are also some other VLP-based heterogeneous positioning systems \cite{yby24, vlp-neuralnet, Lang2021hospital}. Wu \textit{et al.} \cite{yby24} proposed an indoor positioning method combining VLP and quick response (QR) code. The initial positioning was realized by loading a QR code image containing LED position information on the LED lamp. The receiver recognized and decoded the image and then positioned the receiver position. Experimental results showed that the proposed method can achieve an average error of 4.0326 cm, which greatly improves the indoor positioning accuracy compared with the original positioning method. Cao \textit{et al.} \cite{vlp-neuralnet} proposed a memory-artificial neural network (M-ANN) in the 3D indoor RSS-VLP system. M-ANN can efficiently search and retrieve the missing data from an offline simulation database to prevent the VLP outage caused by the blocked LED. The average positioning error of 1.04 cm was experimentally achieved in a test region of 0.6 m $\times$ 0.6 m $\times$ 0.8 m. 
In addition, an RSS-based integrating VLC and power-line communication (PLC) indoor positioning system \cite{Lang2021hospital} was developed for smart hospital applications. The new indoor tracking system consisted of host LED bulbs, user-end optical tags, an LED triangular positioning algorithm, and a PLC interface. Both the simulation and experimental results verified that the proposed hybrid system can achieve centimeter positioning accuracy.

\begin{table*}[htbp]
  \centering
  \caption{Heteroneneous positioning systems}
  \resizebox{\textwidth}{45.3mm}{
    \begin{tabular}{lccccccccccccc}
    \hline
    \multicolumn{2}{c}{\multirow{2}[2]{*}{\textbf{System}}} & \multicolumn{2}{c}{\multirow{2}[2]{*}{\textbf{Positioning Algorithm}}} & \multicolumn{2}{c}{\multirow{2}[2]{*}{\textbf{Experiment/Simulation}}} & \multicolumn{2}{c}{\multirow{2}[2]{*}{\textbf{Auxiliary devices}}} & \multicolumn{2}{c}{\multirow{2}[2]{*}{\textbf{Satial Scale}}} & \multicolumn{2}{c}{\multirow{2}[2]{*}{\textbf{Accuracy}}} & \multicolumn{2}{c}{\multirow{2}[2]{*}{\textbf{Cost}}} \bigstrut[t]\\
    \multicolumn{2}{c}{} & \multicolumn{2}{c}{} & \multicolumn{2}{c}{} & \multicolumn{2}{c}{} & \multicolumn{2}{c}{} & \multicolumn{2}{c}{} & \multicolumn{2}{c}{} \bigstrut[b]\\
    \hline
    \hline
    \multicolumn{2}{c}{\cite{yby2}} & \multicolumn{2}{c}{RSS+Spring Model(Bluetooth)} & \multicolumn{2}{c}{Simulation} & \multicolumn{2}{c}{Bluetooth-beacon} & \multicolumn{2}{c}{5×5×3m} & \multicolumn{2}{c}{6.0 cm} & \multicolumn{2}{c}{High} \bigstrut\\
    \hline
    \multicolumn{2}{c}{\cite{yby3}} & \multicolumn{2}{c}{TOA+Asonic Ranging} & \multicolumn{2}{c}{Experiment} & \multicolumn{2}{c}{Sonic sensor} & \multicolumn{2}{c}{0.5×2.5×1.5m} & \multicolumn{2}{c}{10$\sim$20cm} & \multicolumn{2}{c}{High} \bigstrut\\
    \hline
    \multicolumn{2}{c}{\cite{yasir2014AOA}} & \multicolumn{2}{c}{RSS} & \multicolumn{2}{c}{Experiment} & \multicolumn{2}{c}{Accelerometer} & \multicolumn{2}{c}{5×5×3m} & \multicolumn{2}{c}{\textless 25cm} & \multicolumn{2}{c}{Low} \bigstrut\\
    \hline
    \multicolumn{2}{c}{\cite{yby6}} & \multicolumn{2}{c}{Nearest Neighbor } & \multicolumn{2}{c}{Experiment} & \multicolumn{2}{c}{Geomagnetic sensor} & \multicolumn{2}{c}{Large} & \multicolumn{2}{c}{High} & \multicolumn{2}{c}{Low} \bigstrut\\
    \hline
    \multicolumn{2}{c}{\cite{yby7}} & \multicolumn{2}{c}{RSS+LIPOS} & \multicolumn{2}{c}{Experiment} & \multicolumn{2}{c}{IMU sensor} & \multicolumn{2}{c}{Large} & \multicolumn{2}{c}{Sub-meter} & \multicolumn{2}{c}{Low} \bigstrut\\
    \hline
    \multicolumn{2}{c}{\cite{yby8}} & \multicolumn{2}{c}{RSS+RMSED} & \multicolumn{2}{c}{Experiment} & \multicolumn{2}{c}{Angle sensor} & \multicolumn{2}{c}{1.2×5×2.5m} & \multicolumn{2}{c}{29.8$\sim$46.3cm} & \multicolumn{2}{c}{Medium} \bigstrut\\
    \hline
    \multicolumn{2}{c}{\cite{yby9}} & \multicolumn{2}{c}{EKF-based visual-inertial method} & \multicolumn{2}{c}{Experiment} & \multicolumn{2}{c}{Visual-inertial sensor} & \multicolumn{2}{c}{5×4×2.3m} & \multicolumn{2}{c}{5cm} & \multicolumn{2}{c}{Medium} \bigstrut\\
    \hline
    \multicolumn{2}{c}{\cite{yby11}} & \multicolumn{2}{c}{VLIP} & \multicolumn{2}{c}{Experiment} & \multicolumn{2}{c}{Visual-inertial sensor} & \multicolumn{2}{c}{1.5×1.5×3m} & \multicolumn{2}{c}{2$\sim$15cm} & \multicolumn{2}{c}{Medium} \bigstrut\\
    \hline
    \multicolumn{2}{c}{\cite{cheng2020singlecircle}} & \multicolumn{2}{c}{Image-seneor method} & \multicolumn{2}{c}{Experiment} & \multicolumn{2}{c}{IMU sensor} & \multicolumn{2}{c}{2.7×1.8×1.75m} & \multicolumn{2}{c}{5.44cm} & \multicolumn{2}{c}{High} \bigstrut\\
    \hline
    \multicolumn{2}{c}{\cite{yby15}} & \multicolumn{2}{c}{RSS+PDR} & \multicolumn{2}{c}{Experiment} & \multicolumn{2}{c}{IMU sensor} & \multicolumn{2}{c}{2.0×2.0m} & \multicolumn{2}{c}{14cm} & \multicolumn{2}{c}{Low} \bigstrut\\
    \hline
    \multicolumn{2}{c}{\cite{yby16}} & \multicolumn{2}{c}{RSS+PDR} & \multicolumn{2}{c}{Experiment} & \multicolumn{2}{c}{IMU sensor} & \multicolumn{2}{c}{2.2×2.2m} & \multicolumn{2}{c}{14.5cm} & \multicolumn{2}{c}{Low} \bigstrut\\
    \hline
    \multicolumn{2}{c}{\cite{yby18}} & \multicolumn{2}{c}{Image-seneor method+PDR} & \multicolumn{2}{c}{Experiment} & \multicolumn{2}{c}{IMU sensor} & \multicolumn{2}{c}{Large} & \multicolumn{2}{c}{Sub-meter} & \multicolumn{2}{c}{Low} \bigstrut\\
    \hline
    \multicolumn{2}{c}{\cite{yby21}} & \multicolumn{2}{c}{VL-GraphSLAM} & \multicolumn{2}{c}{Experiment} & \multicolumn{2}{c}{IMU +WIFI sensor} & \multicolumn{2}{c}{Large} & \multicolumn{2}{c}{0.9m} & \multicolumn{2}{c}{High} \bigstrut\\
    \hline
    \multicolumn{2}{c}{\cite{yby23}} & \multicolumn{2}{c}{Image-seneor method} & \multicolumn{2}{c}{Experiment} & \multicolumn{2}{c}{ROS } & \multicolumn{2}{c}{1×1×1.5m} & \multicolumn{2}{c}{0.82cm} & \multicolumn{2}{c}{High} \bigstrut\\
    \hline
    \multicolumn{2}{c}{\cite{yby24}} & \multicolumn{2}{c}{Image-seneor method} & \multicolumn{2}{c}{Experiment} & \multicolumn{2}{c}{QR code} & \multicolumn{2}{c}{2×2×3m} & \multicolumn{2}{c}{4.0326cm} & \multicolumn{2}{c}{High} \bigstrut\\
    \hline
    \end{tabular}%
}
  \label{tab:hete}%
\end{table*}%

\subsection{Summary and Lessons Learned}
We summarize the above articles that are based on the heterogeneous positioning system in Table \ref{tab:hete}. We mainly compare the accuracy, system cost, and the need for auxiliary devices. The emergence of heterogeneous positioning systems is to achieve better performance for LED-based positioning systems. 

However, while having the benefits, heterogeneous positioning systems also face several challenges. For instance, the cost of the heterogeneous positioning systems remains an issue since the fusion of multiple positioning methods means more devices are required at both the transmitter and receiver ends. The use of auxiliary devices will also bring different levels of complexity and implementation difficulties. For instance, the fusion positioning system with the combination of Bluetooth beacons requires additional information processing and beacon placement of Bluetooth. 
In addition, significant variations among different systems present a challenge in integrating multimodal data from various sensors and technologies to achieve more accurate and reliable positioning.
Therefore, in addition to positioning performance, the design of future heterogeneous positioning algorithms should also consider positioning cost and practicality.

To tackle the above challenges, we suggest some countermeasures below. 
First, when integrating VLP with other positioning technologies, the selection of positioning algorithms should balance accuracy, cost, and complexity. 
When using the advantages of different systems to complement each other and improve accuracy, it is crucial to consider the simplicity of the integration implementation. In this regard, the fusion systems of VLP and INS \cite{yby9,cheng2020singlecircle,yby18,wen2023enhanced, alcazar2024seamless} have drawn a lot of attention due to the high degree of sensor integration in intelligent terminals, which offers significant implementation advantages. The fusion system of VLP and visual positioning was proposed \cite{Guo2023graph}, which emphasized the symbiotic advantage that they both use the image sensor for positioning. 
In addition, it is necessary to develop fusion strategies and real-time data processing to improve positioning efficiency. Fusion algorithms such as Kalman filtering and particle filtering are essential for effectively processing and integrating multimodal data from various sources. 
Leveraging machine learning and artificial intelligence is also crucial for optimizing the fusion strategies of sensor data. These technologies enable the system to learn and adjust in real-time, thereby improving the system's adaptability to dynamic environmental changes. 

\section{VLP Network Design}
\begin{figure}
    \centering
    \includegraphics[height=6.5cm]{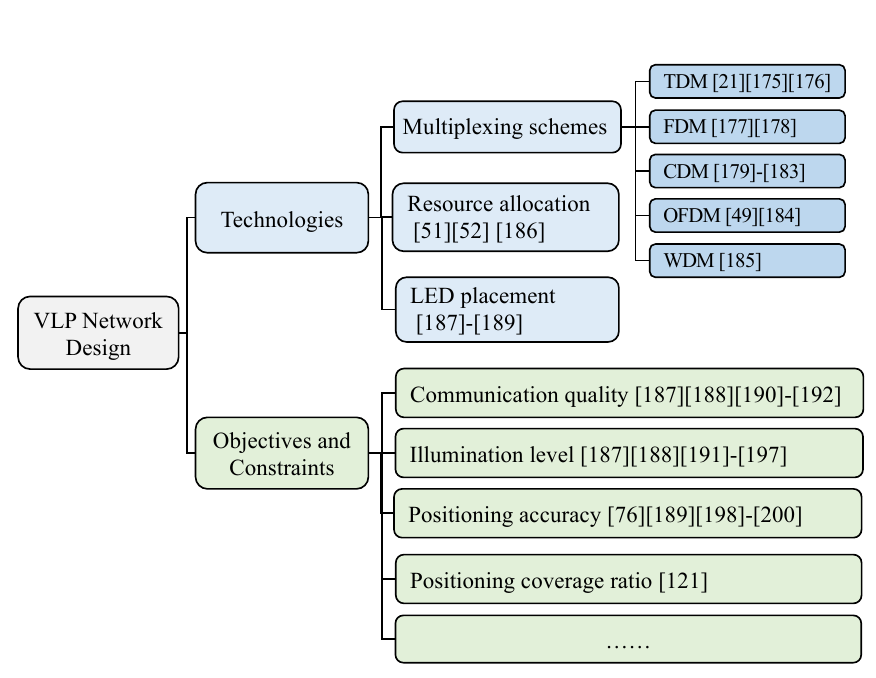}
    \caption{Taxonomy of VLP network design.}
    \label{fig:network}
\end{figure}

In this section, we survey the network design of VLP systems, and Fig. \ref{fig:network} shows the taxonomy. First, we review the technologies in the VLP network design, including multiplexing protocols, resource allocation, and LED placement. Then, we provide some possible objectives and constraints for designing the VLP network. 

\subsection{Technologies in VLP Network Design}
\subsubsection{Multiplexing schemes}\label{multiplexing}
VLP algorithms such as RSS typically require simultaneously receiving multiple visible light signals for positioning. Therefore, appropriate multiplexing schemes are needed to ensure that signals from different LEDs can be differentiated. Here, we summarize the state-of-the-art multiplexing as follows:

\textbf{Time Division Multiplexing (TDM):} In TDM, each lamp is assigned to a time slot and transmits its unique code sequence in its time slot. In the remaining time slots, the LEDs keep illuminating without transmitting any information. Therefore, TDM requires strict time synchronization among LEDs. Zhou \textit{et al.} \cite{zhou2012indoor} developed a TDM method, in which each time was encoded into three 16-bit location codes representing the $x$, $y$, and $z$ coordinates of the LEDs. To keep the transmission power stable, the TDM slot structure was also specially designed so as to provide stable illuminations. However, since each lamp must take up a time slot, the positioning delay can be large when there are a large number of LED lamps, which can limit the application of VLP in mobile scenarios. To tackle this challenge, Hou \textit{et al.} \cite{hou2015multiple} proposed a novel block encoding TDM scheme, in which nine LEDs composed one block to reduce the system delay. Meanwhile, in order to reduce inter-cell interference, an extended binary coded decimal code was used to encode signals transmitted by the LEDs. In addition, two TDM schemes \cite{ho2016uncoordinated} were proposed for VLP systems with/without communication capability, and the required total time slots for estimating channel gains are studied to facilitate positioning.

\textbf{Frequency Division Multiplexing (FDM):} In FDM, each lamp is assigned to a carrier frequency and the transmitted information will be modulated on this frequency. With the assigned carrier frequencies, the LEDs can send information continuously. 
At the receiver side, light signals from different LEDs can be distinguished by filters. Kim \textit{et al.} \cite{kim2012indoor} proposed a carrier allocation
VLC system to mitigate inter-cell interference. At the transmitter side, the QPSK signals with 2 MHz, 2.5 MHz and 3 MHz carriers are generated. The receiver position can be calculated by the trilateration method. For instance, a VLP system based on FDM method was proposed \cite{de2015visible}, which utilized the properties of square waves in the frequency domain. Neighboring LEDs used multiples of the ground frequency of the first LED, and the receiver performed an FFT on the received signals to obtain the RSS values. The system can utilize unsynchronized low-bandwidth transmitters, thus achieving an easy implementation with current high-efficiency LED drivers.

\textbf{Code Division Multiplexing (CDM):} In CDM, each lamp is assigned to a unique code and these codes are orthogonal to each other. The LEDs use digital modulation to send information, and the receiver utilizes the same codes to obtain different signals. 
The work \cite{de2014optical} used the CDM scheme for the VLC system and evaluated uni- and bi- polar codes with the appropriated receiver. The results showed that
bipolar codes can reduce the distance error caused by surrounding light. Then, Szabo \textit{et al.} \cite{szabo2015investigation} designed and built an experimental platform for the VLC CDM system, and verified the functionality of the system. 
Ho \textit{et al.} \cite{ho2017coding} formulated a nonconvex optimization problem to find the optimal code and used majorization theory analytically to solve this optimization problem. 
It has been proved that a combinational code is an optimal code that simultaneously attains the minimum values of both total noise variance and maximum noise variance. Through the optimal code, a bound on the fundamental trade-off between the total noise variance and codeword length is derived. 
Saed \textit{et al.} \cite{saed2018combinational} used combinational code as a coding scheme for channel estimation, and the experimental results showed that combinational code significantly outperforms the other two schemes based on TDM in terms of noise variance. Another work \cite{aparicio2019visible} proposed an encoding scheme based on 1023-bit Kasami sequences for every transmission, which provided multiple access capability and robustness against low signal-to-noise ratios and harsh conditions, such as multipath and near-far effect.

\textbf{Orthogonal Frequency Division Multiplexing (OFDM):} In OFDM, the transmission bandwidth is divided into a series of orthogonal subcarriers, and different lamps are assigned to different subcarriers.
There is no need for bandwidth protection between subcarriers, and the spectrums can overlap with each other. Therefore, compared to FDM, OFDM can fully utilize the bandwidth resources. An indoor VLP and VLP system using the OPDM scheme was proposed \cite{lin2016indoor, lin2017experimental} to overcome intercell interference and provide indoor positioning and data communications. The transmitted signals from LEDs were encoded with an allocated subcarrier, and the receiver recovered all signals by a DFT operation. Three subcarriers with the maximum RSS are selected for positioning based on the trilateration method. The experiment results showed that the proposed system can achieve a mean positioning error of 1.68 cm and an error vector magnitude of more than 15 dB.

\textbf{Wavelength Division Multiplexing (WDM):} In WDM, different lamps send optical carrier signals with different wavelengths to transmit information. At the receiver side, optical carriers with different wavelengths are separated to restore the original signals. A localization method named color-based localization \cite{pergoloni2017visible} was proposed, which utilized the properties of metamerism. The red-green-blue LEDs provided a white light sensation to the human eye, and the CSK scheme allowed data transmission by modulating the intensity of LEDs using a color coding format. The receiver position can be estimated using the RSS or TDOA method.

\subsubsection{Resource allocation}

VLP is based on VLC since VLP needs to know the location information of each LED, which is transmitted by VLC. Conversely, the VLC channel model is a function of the receiver’s location. They interact to achieve an efficient integrated communication and positioning system. 
In practical indoor scenarios, both communication and positioning services should be required simultaneously. Therefore, reasonable resource allocation provides an important guarantee for high-speed communication and high-accuracy positioning. 

There have been some works focused on this issue. 
For instance, Ma \textit{et al.} \cite{ma2022optimalpowerallocation} revealed the intrinsic relationship between VLP and VLC in a single-LED-based VLCP system, and it demonstrated that the positioning information can be used to estimate the channel state information. A robust power allocation scheme was proposed under the practical optical constraints and QoS requirements. The result showed the efficiency in both localization and communications.
Then, Yang \textit{et al.}\cite{yang2019qos} presented an integrated VLCP network for indoor IoT devices and proposed jointly optimizing the access point selection, subcarrier group allocation, adaptive modulation, and power allocation approach to satisfy different QoS requirements of devices while maximizing the network data rate. They further presented an integrated visible light VLCP system and proposed a jointly coordinated subcarrier and power allocation approach with the purpose of maximizing the system sum rate while guaranteeing the minimum data rate and positioning accuracy requirements of devices \cite{yang2020coordinated}. 

\subsubsection{LED Placement}
In VLCP systems, the LED has a double function of illumination and communication. 
The LED layout can directly affect illumination uniformity, communication performance, and positioning performance. 
Therefore, the layout of the LEDs must be properly designed to make the illumination uniform while ensuring the communication quality for positioning. 
To this end, the work \cite{yang2020placement} proposed a power-efficient LED placement optimization algorithm for the VLC system, which considered the illumination level and communication requirements simultaneously. The simulation results show that the proposed LED placement algorithm can harvest a 14\% power consumption gain. Li \textit{et al.} \cite{Li2022deployment} also studied the optimal deployment of LEDs indoors under the constraints of illumination and signal-to-noise ratio (SNR) by using simplicial complexes. It achieved seamless coverage with the minimal number of LEDs required. 

Based on the VLC systems, when designing the LED placement for VLP systems, the positioning performance such as accuracy and coverage ratio, should be considered in addition to communication performance metrics. To evaluate the positioning performance, a 3D normalized positioning error metric (NPEM) \cite{xu2022indoor} was characterized to explore the relationship between the NPEM and the LED cell layout. Then, the authors proposed a layout optimization algorithm for a structured square LED transmitter cell to minimize the NPEM. 
Limited attention has been directed towards the LED placement problem in VLP systems. The primary challenge lies in formulating a theoretical model that correlates positioning performance metrics, like accuracy, with the arrangement of LEDs. Consequently, it is meaningful to create a theoretical model for positioning accuracy that holds significant value. Such research can serve as a foundational framework for enhancing performance.

\subsection{Objectives and Constraints in VLP Network Design}
This section presents the possible objectives and constraints in VLP network designs. Since VLP and VLC systems are homogeneous, in addition to the objectives and constraints in VLC systems, the VLP network should also consider some other unique challenges. We list several possible objectives and constraints as follows. 
\subsubsection{Communication Quality}
The communication quality of the VLP system is the basis to ensure the positioning function. Different from traditional radio frequency systems,  the common Shannon channel capacity formula cannot be applied to VLC. 
The illumination demands and the LED dynamic range constraints must be considered when deriving channel capacity, and the VLC channel capacity can be expressed as \cite{wang2013tight}  
\begin{equation}\label{eq:commun_capacity}
    C \ge \frac{1}{2}{\log _2}\left( {1 + \frac{2}{{2\pi }}{{\left( {\frac{{\xi Ph}}{\sigma }} \right)}^2}} \right)
\end{equation}
where $C$ is the channel capacity, and $\xi$ represents the illumination target. $P$ is the optical power of the LED, and $\sigma$ is the standard deviation of the additive white Gaussian noise. 
The expression of (\ref{eq:commun_capacity}) has been adopted in optimizing the network in some VLC systems \cite{yang2020placement, yang2019UAV}. 
In addition to the above channel capacity, several studies also used SNR to quantify communication quality \cite{Li2022deployment, mushfique2020optimization}.

\subsubsection{Illumination Level}
VLP systems reuse the luminaires in the room, and thus, the illumination level should be taken into consideration. 
The brightness is typically defined by the luminous intensity. The illumination level constraints can be determined by the minimum illumination or average illumination in the room. 
The illumination in lux, ${\eta _{j}}$, at the receiver point $j$ can be given by 
\begin{equation}\label{eq:illumilevel}
    {\eta _{j}} = I_0{\cos ^m}({\phi _{j}})\cos ({\psi _{j}})/d_{j}^2
\end{equation}
where ${I_0} = I\left( {\Phi  = 0} \right) = \left( {m + 1} \right){{\phi _{j}}}/\left({{2\pi }}\right)$ is the maximal luminous intensity. $\Phi  = 683\int_0^\infty  {V\left( \lambda  \right)} P\left( \lambda  \right)d\lambda $ is the luminous flux, where $V\left(\lambda \right)$ is the luminosity function and $V\left(\lambda \right)$ is the spectral power distribution \cite{li2018optimization}.

To satisfy the illumination requirements, some studies \cite{Li2022deployment, li2021led, yang2019UAV} set a threshold for the illumination function (\ref{eq:illumilevel}) of each sample point at the receiver plane to formulate a network optimization problem. In addition, some studies \cite{yang2020placement, yang2021joint, wang2022cvrmse, mushfique2020optimization} ensured the illumination level from a perspective of illumination uniformity. For instance, the coefficient of variation of root mean square error (CV(RMSE)) was typically adopted \cite{yang2020placement, yang2021joint, wang2022cvrmse}, and CV(RMSE) can be expressed as 
\begin{equation}
    {\rm{CV}}\left( {{\rm{RMSE}}} \right) = \left( \sqrt {\frac{1}{N}\sum\limits_{j = 1}^N {{{\left( {{\eta _{j}} - {\eta _{{\rm{avg}}}}} \right)}^2}} } \right) /{\eta _{{\rm{avg}}}},
\end{equation}
where ${\eta _{{\rm{avg}}}}={\frac{1}{N}}{\sum\limits_{j = 1}^N {{{ {{\eta _{j}} }}} }}$ is the average illumination of all sample points. 
There is also the definition of illumination uniformity that is given by the ratio between the minimum and the average illumination intensity among all sample points, i.e., $\vartheta  = \min \left( {{\eta _j}} \right)/{\eta _{{\rm{avg}}}}$ \cite{standardization2002lighting}, which has been adopted to achieve acceptable illumination uniformity in the room \cite{mushfique2020optimization}.

\subsubsection{Positioning Accuracy}
In contrast to a singular VLC network, the precision of positioning is intricately linked to VLP networks. As highlighted in Section \ref{seq:homoVLP}, the efficacy of positioning algorithms is markedly impacted by the count of received LEDs, a factor confined by the arrangement of LEDs. Few studies delve into network optimization strategies that account for VLP accuracy. The Cramer-Rao lower bound (CRLB) is a common tool employed to assess the theoretical positioning accuracy of VLP.
The location and orientation of the user are bounded from below as follows \cite{zhou2019performance}
\begin{equation}
    \left\{ \begin{array}{l}
{\mathbb{E}}\left\{ {\left\| {{\bf{\hat x}} - {\bf{x}}} \right\|_2^2} \right\} \ge {\rm{tr}}\left( {{{\rm \mathcal{B}}_{\bf{x}}}\left( {{\bf{x}},{\bf{u}}} \right)} \right)\\
{\mathbb{E}}\left\{ {\left\| {{\bf{\hat u}} - {\bf{u}}} \right\|_2^2} \right\} \ge {\rm{tr}}\left( {{{\rm \mathcal{B}}_{\bf{u}}}\left( {{\bf{x}},{\bf{u}}} \right)} \right)
\end{array} \right. 
\end{equation}
where $\mathbb{E} \left\{ \cdot \right\}$ is the expectation over noise. ${\bf{\hat x}}$ and ${\bf{\hat u}}$ are the estimation of the location and orientation, respectively, while $\rm{tr} \left\{ \cdot \right\}$ represents the matrix trace. ${{\mathcal{B}_{\bf{u}}}\left( {{\bf{x}},{\bf{u}}} \right)}$ and ${{{\mathcal{B}}_{\bf{x}}}\left( {{\bf{x}},{\bf{u}}} \right)}$ denote the CRLBs on the location estimation error and orientation error, respectively, and they are given by 
\begin{equation}
    \left\{ \begin{array}{l}
{{\mathcal{B}}_{\bf{x}}}\left( {{\bf{x}},{\bf{u}}} \right) = \left( {{\mathcal{H}}_{\bf{x}}^{{\rm{obs}}} + {\bf{X}_{{\rm{prior}}}}} \right) ^{-1}\\
{{\mathcal{B}}_{\bf{u}}}\left( {{\bf{x}},{\bf{u}}} \right) = \left( {{\mathcal{H}}_{\bf{u}}^{{\rm{obs}}} + {\bf{U}_{{\rm{prior}}}}} \right) ^{-1}
\end{array} \right. 
\end{equation}
where ${\mathcal{H}}_{\bf{x}}^{\rm{obs}}$ and ${\mathcal{H}}_{\bf{u}}^{\rm{obs}} $ denote the observation information of the location and orientation of the user, respectively, and ${\bf{U}_{{\rm{prior}}}}$ and ${\bf{X}_{{\rm{prior}}}}$ denote the prior information. 
The obtained CRLBs were then used to comprehensively study the performance limits of RSS-based VLP algorithms. The authors also pointed out that the CRLB could be used to provide guidelines for optimizing the performance of practical VLP systems. Then, Shi \textit{et al.} \cite{shi2023joint} optimized the energy efficiency performance for a radio frequency-visible light communication and positioning system by guaranteeing the CRLB of the positioning error and the minimum data rate of the communication. Ma \textit{et al.}  \cite{ma2023waveform} derived CRLB to quantify the performance of the integrated VLP and VLP system, and aimed to minimize the CRLB of VLP, i.e., the positioning errors, while satisfying the outage probability constraint of the VLC rate, and total power constraint. 

In addition to CRLB, Zhang \textit{et al.} \cite{zhang2022beacon} derived a novel closed-form error expression that related the noise statistics to the LED localization error at a certain point and provided insights into the error control of the LED coordinates databases.

However, it is still challenging to construct a theoretical accuracy model for camera-based VLP systems.
Recently, Xu \textit{et al.} \cite{xu2022indoor}  introduced a novel positioning error metric, NPEM, derived from the partial derivative of the positioning function. Their analysis revealed an inversely proportional relationship between NPEM and the number of captured LEDs. Intriguingly, NPEM exhibited independence from the rotation angle around the z-axis when the transmitter and receiver planes were parallel. This finding serves as a foundational concept for devising a quantitative model for positioning errors by amalgamating positioning functions from distinct camera-based VLP systems.

\subsubsection{Positioning Coverage Ratio}
When considering the performance of VLP, the coverage ratio of the positioning service should also be considered in the VLP network. 
As shown in (\ref{eq:cover}), the coverage ratio is the proportion of the effective area that the positioning service can be provided. 
In the VLP network design, eq. (\ref{eq:cover}) can be formulated as an optimization objective, or as a constraint to guarantee the positioning service area. For instance, let $A_{\rm{affective}}$ be the area where the user can achieve a positioning accuracy of 5 cm, and then, we can have
\begin{equation}
    {R_{{\rm{cov}}}} = \frac{{{A_{{\rm{effective}}}}}}{{{A_{{\rm{total}}}}}} \times 100\%  \ge 95\%
\end{equation}
with (\ref{eq:cover}) to guarantee the accuracy and coverage of the VLP network system simultaneously.

In addition, there are other objectives that can be conducted in the VLP network system design, such as maximizing the system efficiency and minimizing the deployed LEDs. In general, it is necessary to make the VLP system more efficient, accurate, and reliable.

\subsection{Summary and Lessons Learned}

In VLP applications, the network design is significant in optimizing the system's efficiency, performance, and reliability in communication and information processing. An effective network design can significantly enhance the system's ability to handle complex data, reduce latency, and improve overall throughput, thereby ensuring robust and reliable VLP service. However, despite its importance, there is relatively limited research specifically focused on the network design of VLP. The unique requirements and complexities of VLP networks, such as the optimal placement of LEDs and receivers and resource allocation, still deserve investigation.

Future research should evaluate positioning systems comprehensively rather than solely positioning accuracy. Achieving a comprehensive positioning system requires jointly considering communication quality, illumination service, position accuracy, coverage, etc. 
In addition, in line with the trend of integrated sensing and communication in 6G \cite{Saad2020vision6G}, the network design within VLCP is also worth investigating. This can enhance the performance and widespread application of VLP systems while providing a more holistic approach to indoor positioning and communication. Such advancements are crucial to meeting the growing demands of smart environments and IoT applications.


\section{Applications Scenarios of VLP}
\begin{figure}
    \centering
    \includegraphics[width=1\linewidth]{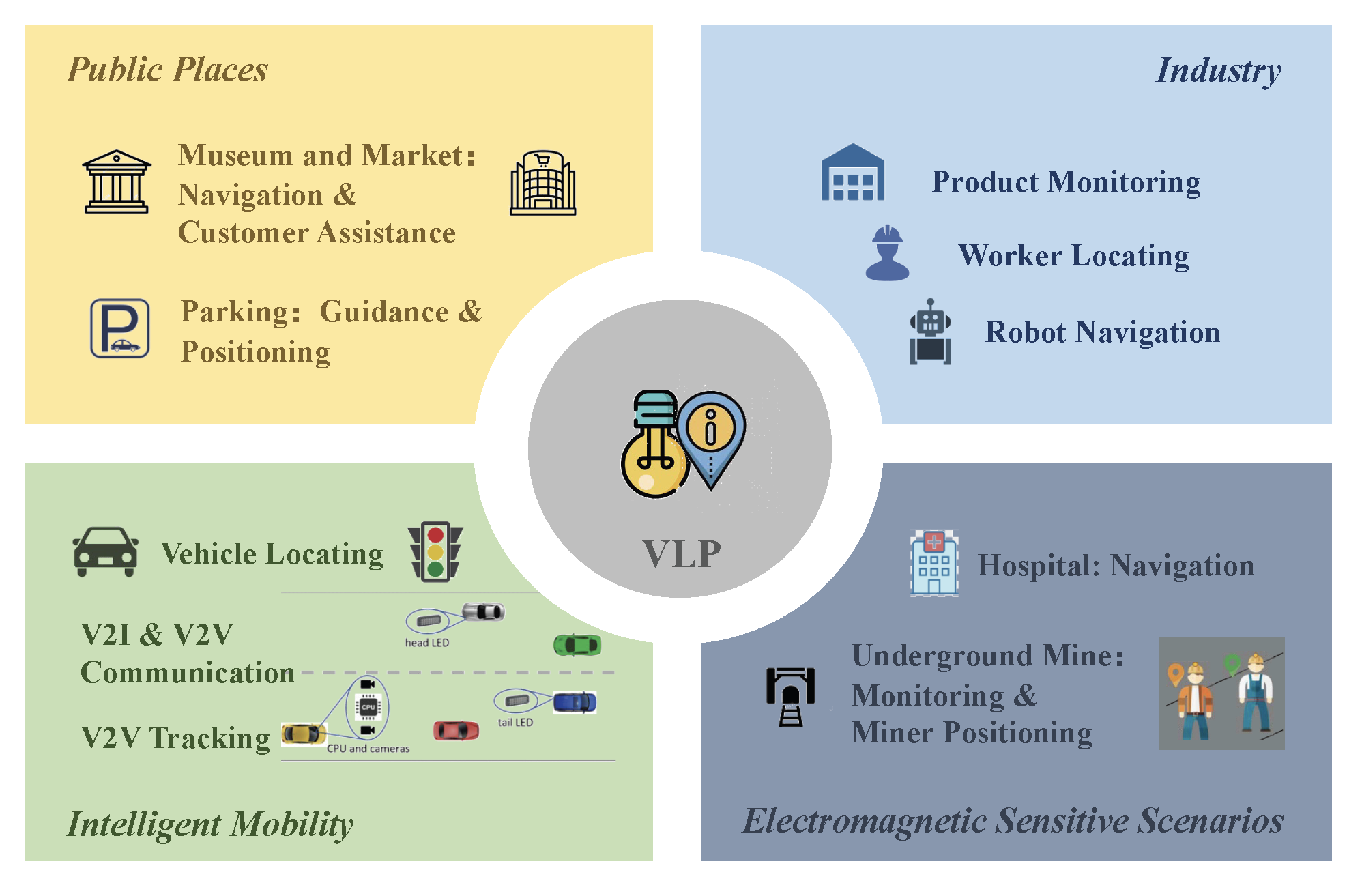}
    \caption{Applications of VLP.}
    \label{fig:application}
\end{figure}

After detailing the technologies in VLP, this section introduces the potential applications of VLP in indoor positioning scenarios. VLP can provide a wide range of indoor positioning services. In particular, in addition to the common indoor positioning scenarios, VLP is especially suitable for electromagnetic-sensitive indoor places. The main applications of VLP indoor positioning services are listed in Fig. \ref{fig:application}.

\subsection{Public Places} 
Location-based services are especially important in large public venues, such as shopping malls, museums, transport stations, parking, and so on. VLP reuses the existing indoor lighting sources, which can conveniently provide services for users. Meanwhile, some information about the venues can also be transmitted to the users by using VLC. 
For instance, the unlicensed visible light of the electromagnetic spectrum was used to provide museum visitors with high-accuracy positioning, multiple forms of interaction services, and high-resolution multimedia delivery on a mobile device \cite{vlp-5g}. 
The visible light positioning technology was also applied to the supermarket \cite{he2022supermarket}. The supermarket LED lighting was used to transmit the commodity information to solve the problem that it is inconvenient to find commodities in large supermarkets.
Moreover, Qiang \textit{et al.} \cite{qiang2019parking} proposed an indoor-parking intelligent navigation system with the use of LED light, which realized real-time detection of parking space status, real-time high-precision positioning, and navigation for vehicles and pedestrians. Madahian \textit{et al.} \cite{madahian2023hybrid} proposed a hybrid VLC and RF positioning system for parking automation system.

\subsection{Industries}
In industries such as warehouses, location-based services are required for worker locating, product monitoring, and robot navigation. 
In 2019, Lam \textit{et al. } \cite{lam2019industry} proposed a ray-surface positioning algorithm by employing a steerable laser in conjunction with a lighting luminaire to position devices. An application of positioning was demonstrated for real-time robot control in an interactive multiparty cyber-physical-virtual deployment. Then, in 2021, Du \textit{et al.} \cite{du2021iot} also proposed an RSS-based VLP position estimator for industrial IoT. The experimental results indicated that using VLP can significantly reduce training time for offline preparation by more than 80\% when compared to that of deep neural networks, support vector machines, and random forest. 

\subsection{Intelligent Mobility}
The headlights, taillights of vehicles, traffic lights, and street light posts enable communication of Vehicle-to-Vehicle (V2V) and Vehicle-to-Infrastructure (V2I) through VLC \cite{mare2016intelligenttransport}. Therefore, some researchers proposed the use of VLP in intelligent mobility. 
Singh \textit{et al.} \cite{singh2022V2I} proposed a positioning scheme for the V2I environment using optical camera communication. The vehicle camera receives the signals from at least three streetlights to estimate its localization. The experimental demonstration verifies that the proposed technique can achieve an accuracy of 1 m.  Do \textit{et al.} \cite{do2019V2V} proposed a V2V tracking system based on VLC, which used head and tail lamps on the vehicle to transmit the positioning signals and used two CMOS cameras to receive the signals. The system used the geometric relationship between the two cameras and the image of the LEDs to locate the target vehicle. In addition, a modified Kalman Filter is used to smooth the tracking result.

\subsection{Electromagnetic Sensitive/Unavailable Scenarios}
In addition to the common large venues, the VLP techniques can also provide location services for electromagnetic-sensitive or electromagnetic-unavailable scenarios. For example, in the hospital, RF wireless is often prohibited because RF is harmful to human beings and can interfere with healthcare
equipment. In such circumstances, VLP can be a potential technique to provide reliable positioning service, which can be used for the navigation of wheelchairs and patients. Visually impaired or handicapped people can also benefit from the high accuracy of VLP systems in avoiding collisions or falls. 
Lang \textit{et al.} \cite{Lang2021hospital} designed an integrated VLP and PLC system for smart hospitals. The system used RSS for positioning and consisted of LED bulbs and user-end optical tags. The proposed system can provide smart dispatching medical personnel, deploying medical instruments, and retrofitting hospital floor layout.  

Moreover, VLP technology shows excellent potential in the underground mine with complex environments like occlusion by ores. In a scenario of underground mines, a Cell-ID positioning technology \cite{nicolas2016mine} was proposed by combining with VLC. The lamps were deployed on the tunnel and assigned a specific frequency. The receiver's location was inferred by judging the frequency information. Then, Pang \textit{et al.} \cite{pang2022undergroundVLP} proposed a novel reverse VLP mechanism to achieve one-dimension underground mine positioning, in which the user's headlamp broadcasted the information while the receivers were mounted on the ceiling. 

{\subsection{Summary and Lessons Learned}

VLP has the potential to enhance the user experience in public spaces like shopping malls and museums through high-precision navigation. It can also help advance the intelligentization in industries with worker and product tracking, and achieve intelligent mobility through V2V and V2I communication.   
Moreover, VLP is advantageous in RF-sensitive or unavailable scenarios, such as hospitals and underground mines, where it provides safe and reliable navigation for patients and workers.

Hence, VLP's integration with existing infrastructure, high accuracy, and reliability make it a promising technology for future indoor positioning applications. However, the practical applications of VLP have progressed slowly. Up to now, only a limited number of companies have engaged in the design and application of VLP products, such as Phillips Lighting \cite{Phillips} and ByteLight \cite{ByteLight}.
In this circumstance, it is necessary to explore efficient business models, so that VLP can attract more attention from the industry to promote the development of VLP applications. 
In addition, the challenges of transitioning VLP from theoretical advantages to wide applications also need to be conquered to advance the development of VLP applications. The next section will detail these challenges.

\section{Challenges and Opportunities}

Research on LED-based indoor VLP systems has explored many techniques to improve positioning performance. However, there are still many challenges in this field. In this section, we highlight some of the significant challenges and suggest the corresponding possibilities.

\subsection{Positioning information} The positioning accuracy is directly related to the received positioning information.
According to the receiver type in VLP, the camera and the PD face different challenges in receiving positioning information.

When PD is used as the receiver, it is sensitive to ICI in multi-cell scenarios, i.e., the room that is deployed multiple LEDs. Different LED cells need to modulate different VLC information for LED ID recognition. When adjacent cells use the same frequency band, it will produce interference. In this circumstance, some consideration should be given to the ICI problem. The capabilities of multiplexing schemes, such as TDM, FDM, CDMA, etc. mentioned in Section \ref{multiplexing}, can be further studied and validated to apply to LED-based VLP systems. Also, the research on appropriate LED layouts may also help reduce the impact of ICI on positioning performance. 
In addition, multipath reflection is also a major factor that influences the accuracy of the received positioning information. The work \cite{gu2016multipath} investigates the impact of multipath reflections on indoor VLP systems. It is shown that the corner area suffers from severe reflections, and the average positioning error here is 1.5 m higher than that of the center area. Although the impact of reflections decreases at the edge area, it also causes a positioning error of nearly 1 m. However, the multipath reflection is often considered as a strong random noise \cite{Hossei2020finger} in the existing works. Hence, it can be seen that the multipath reflection should be eliminated to ensure the accuracy of VLP systems.

When the camera is equipped as the receiver, it typically uses the rolling shutter effect of the CMOS camera to capture fringe images of LEDs to decode the VLC information. The camera, as the receiver, can get rid of ICI and multipath reflections, but the fringe images can affect the extraction of the LEDs' visual information. 
It is known that some works \cite{bai2021vp4l, cheng2020singlecircle} have considered using the shapes of the luminaire for positioning by using the camera as the receiver. However, they did not consider how fringe images affect the accuracy of visual information extraction, which can influence positioning performance. 
In addition, the camera may fail to receive the VLC information if the size of the luminaire is too small or the communication distance is large. This is because the rolling shutter effect of a CMOS camera is unavailable in this circumstance. 
Therefore, a stable VLC system deserves investigation for receiving the positioning information accurately for the VLP system.

\subsection{Reliability} For PD-based VLP, the positioning reliability relies on accurate channel model estimation. The commonly used channel models, such as the Lambertian model, require the LED lamp to be Lambertian light source, which may not be true in practice due to the type of source, the LED masks, dust, ambient light, and other possible factors \cite{yang2021CARSS}. Hence, it is still challenging to realize highly reliable VLP algorithms under practical non-ideal VLC channels. 
On the other hand, when using a camera as the receiver, the information is extracted from the captured images. The performance of the camera, occlusion, and other environmental factors can also influence positioning reliability. For instance, the ISO value of photography represents sensitivity to light and has a major effect on the images, and thus it directly decides whether the fringe images can be captured to extract the VLC information. In addition, it is more prone to image noise in dark indoor environments, which can also lead to positioning unreliability. 
Another important factor affecting the reliability of VLP is ambient occlusion. Regardless of whether the PD or the camera is used as a receiver, occlusion will cause positioning failure. Hence, some anti-occlusion VLP algorithms should be investigated to enhance positioning reliability. 
Changes in the environment, such as moving objects or varying light conditions, can also affect the reliability of VLP. 

In this circumstance, continuous monitoring and adaptive algorithms are required to maintain accurate positioning even in dynamic scenarios.
In fact, machine learning, particularly deep learning~\cite{9562559,9210812,chen2021communication} can be a potential to address the reliability issues in VLP systems, such as interference from ambient light and object occlusions. 
The attempts of machine learning in the existing VLP systems \cite{zhang2019high, saboundji2022accurate, aparicio2022experimental, tran2022machine, yuan2018tilt, liu2022machine, alenezi2022machine} demonstrate the potential that leveraging machine learning improves the performance of VLP systems. 



\subsection{Cost} Although the low cost is one of the advantages of the LED-based VLP system, the cost issue is still one of the challenges limiting the practical application of VLP at the present stage. 
The cost in VLP primarily lies in the deployment and maintenance of the infrastructure required for accurate positioning. 
In particular, VLP relies on specialized LED lights with positioning capabilities, which require retrofitting the existing commercial LEDs. 
Presently, the LEDs utilized for VLP are yet to undergo industrial-scale production, and their individual coverage remains limited. Considering large-scale deployments in indoor settings, these components can incur significant costs. Moreover, integrating an LED-based VLP system might necessitate merging with existing frameworks like building management systems or mobile applications. The expenses related to adapting or upgrading these systems to accommodate visible light positioning can further contribute to the overall expenses.

In essence, the positioning algorithms need to be innovated to reduce dependence on custom LEDs, thereby leveraging fewer custom LEDs and even unmodified LED infrastructures. For instance, Litell \cite{zhang2016litell} and NaviLight \cite{zhao2017navilight} were both proposed for positioning based on unmodified light sources, which relaxes the limitation on custom LED hardware. 
In addition, developing economical LED modulation hardware tailored for VLP necessitates further research. 
VLP technology relies on custom LEDs capable of both illumination and precise positioning. However, current commercial LEDs were primarily designed for lighting purposes and do not inherently support the sophisticated modulation and communication required for high-precision VLP systems. 
This gap necessitates the advancements of new LED hardware, coupled with their industrialization, to lower the application costs of VLP. Nonetheless, as technology progresses and economies of scale come into play, it is anticipated that the costs associated with VLP systems will gradually decrease over time.


\subsection{Industrialization} The industrialization of VLP is in its early stages, and VLP is not yet widely adopted. Several companies are actively working on commercializing VLP systems. 
For instance, as early as 2017, Philips Lighting installed the first supermarket with Philips’ indoor positioning system in Germany. His system enabled location-based services to identify items within a 2,400 $\rm{m}^2$ shop area with an accuracy of 30cm. 
Another company, ByteLight, established by Boston University alumni, specializes in real-time indoor location and movement tracking using LED lighting. Their focus lies in creating a mobile-based services platform catering to retail, commercial, and corporate enterprise customers. ByteLight aims to collaborate with forward-thinking LED lighting manufacturers to embed their technology directly into lighting products, facilitating market availability.
In addition, the company, Interact, is implementing an indoor navigation partner program to unlock the full potential of its indoor positioning system. 

It can be seen that VLP is still an emerging technology. There is a lack of principles including standardized protocols and frameworks for implementation, which makes it difficult to develop interoperable and scalable positioning systems, thus hindering the industrialization process. 
Meanwhile, the emergency and development of heterogeneous positioning systems provide new possibilities for the application of VLP but bring new challenges to formulating such unified rules.
Hence, a set of consensus VLP system principles is crucial \cite{Do2016An} for fostering academic research and industrialization of VLP. 
In addition to principles, the industrialization of VLP is also closely related to hardware innovation. This necessitates the development of new LED hardware that can promote the industrialization of VLP.

\subsection{Universality} LED-based VLP systems currently do not have the ability to be applied across a wide range of scenarios, environments, and use cases. 
In diverse environments, there are usually different scales of LED layouts and different LED models. Hence, achieving the seamless integration of positioning while maintaining positioning performance can be a hurdle, which is technically challenging and costly. In addition, compatibility issues could arise when dealing with a variety of receiver devices, such as smartphones, tablets, wearables, and other devices. Hence, compatibility with receiver devices is another issue that restricts the universality of LED-based VLP systems.

To address these hurdles, a plausible approach involves modifying a segment of the deployed indoor LEDs when integrating the VLP system into the existing lighting infrastructure. This targeted modification minimizes technical costs. Consequently, investigating efficient LED selection and layout strategies becomes crucial. Thus, efficient LED selection and layout schemes illustrated in Section V are worth investigating. 
Moreover, reducing the number of required LEDs also contributes to enhancing the universality of VLP systems since it can enhance the flexibility of VLP in environments with varying architectural designs and lighting layouts. Some researchers have focused on this point and proposed single LED-based VLP algorithms \cite{zhang2017single,cheng2020singlecircle,zhu2024visible}. 
In addition, to address the compatibility issue, it is crucial to study standardized protocols to ensure that various receivers can seamlessly access and utilize positioning signals from LEDs. Hardware adaption is also necessary since integrating VLP systems with existing devices can reduce the need for significant modifications so that enhancing the university of VLP. 
This necessitates using ubiquitous and simple receiver devices for positioning, avoiding using expensive lenses or special PD arrays. 
For instance, several studies \cite{guan2019image, bai2021vp4l} proposed using the smartphone's built-in camera to implement VLP systems. In addition, other studies \cite{cheng2020singlecircle, wang2021arbitrarily} proposed leveraging both the camera and IMU for positioning, both of which are ubiquitous in modern intelligent devices.

\subsection{Privacy and Security}

Human and device location information is considered sensitive data that can expose users to various risks including stalking, theft, and even security threats. 
Although the high precision of VLP systems is advantageous, it may also raise significant privacy concerns when the location data is without consent.
In addition, VLP signals are typically confined to illuminated areas, and thus they can potentially be intercepted by unauthorized receivers if not properly secured. Due to the sensitivity to occlusion and ambient light, VLP signals can be disrupted or manipulated, leading to inaccurate positioning data. 
However, security and privacy issues in VLP have not garnered as much focus as those in the field of communications. 
Since the receiver in VLP systems often operates within strict energy and computation constraints, complex methods are not expected to ensure the privacy and security of location data. 
Moreover, with the development of heterogeneous positioning technologies, the receiver may use diverse technologies and information, and each of them has its own vulnerabilities and security implications. This diversity complicates the tasks of a universal solution for security and privacy. 

From a technological perspective, enhancing location data security requires a multi-faceted approach. This could involve the development of lightweight cryptographic algorithms suitable for energy-constrained devices, advanced anonymization techniques to protect user identities, and robust access control mechanisms. Moreover, standardized security protocols should also be considered to ensure a cohesive and secure framework. 
By addressing these challenges, it is possible to foster trust and promote broader adoption of VLP applications, balancing the precise location services with the imperative of protecting individual privacy and security.

\section{Conclusion}
The increasing demand for location-aware services has propelled the rapid development of LED-based VLP technology. Acknowledged for its promising capabilities, VLP has garnered recognition from both academia and industry. 
This paper offers a comprehensive survey of VLP systems. As the communication basis of VLP, the VLC technique was overviewed first.
Then, we categorized the existing LED-based positioning systems into homogeneous VLP systems and heterogeneous positioning systems, and we overviewed the positioning algorithms belonging to the two types.
The positioning accuracy and the coverage performance were also compared. 
Furthermore, this survey unveils the prevalent challenges and opportunities within this burgeoning field.
Ultimately, our aim is to furnish researchers with an accessible resource that encapsulates the breadth of work in this domain, enabling them to identify valuable research avenues within the realm of LED-based VLP.

\bibliographystyle{IEEEtran}
\bibliography{main_tex_final.bib}

\vspace{-3ex}
\begin{IEEEbiography}[{\includegraphics[width=1in,height=1.25in,clip,keepaspectratio]{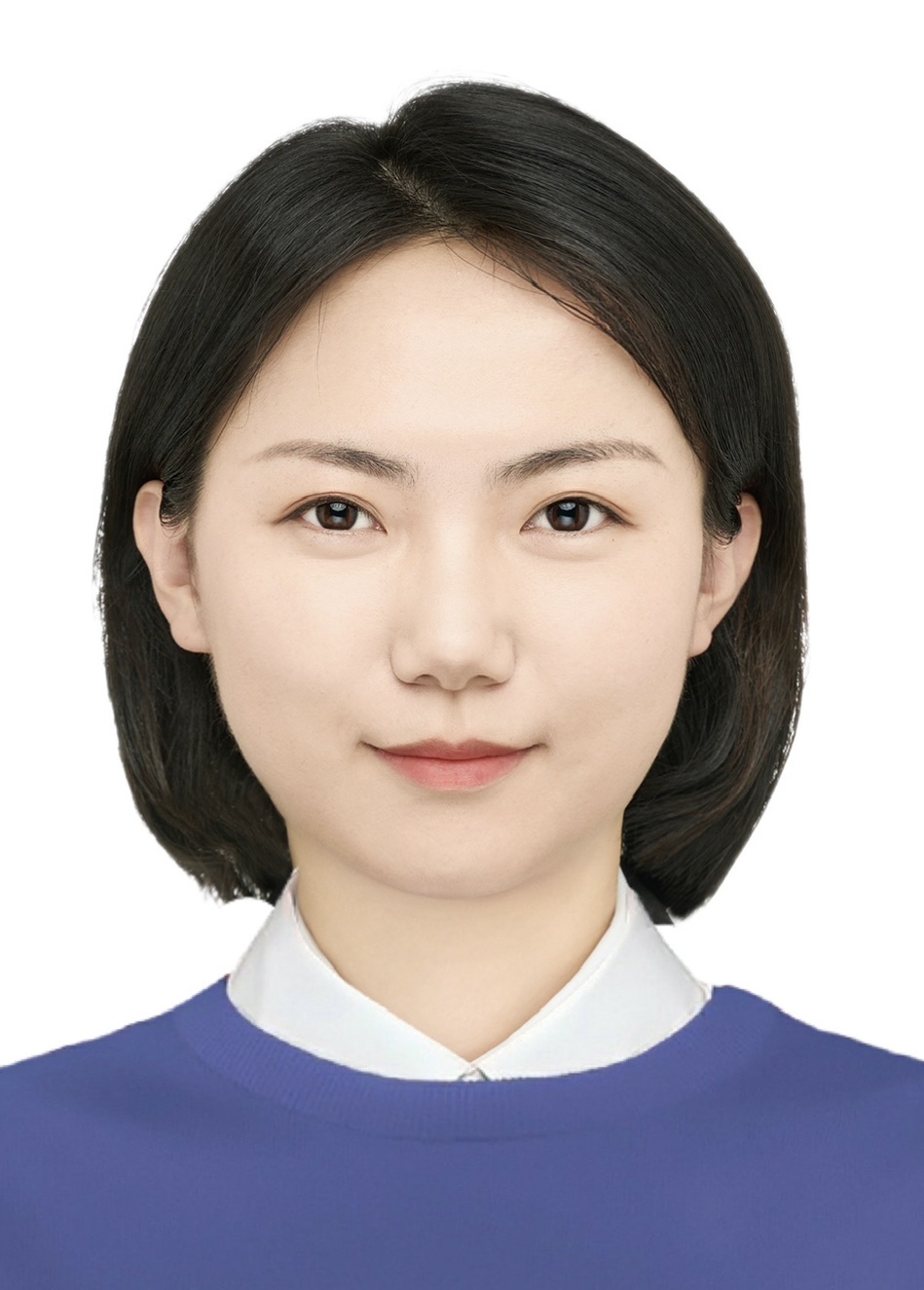}}]{Zhiyu Zhu} received the B.S. degree in information engineering from Southwest Jiaotong University (SWJTU), Chengdu, China, in 2018, and the Ph.D degree in Information and Communication Engineering from School of Information and Communication Engineering, Beijing University of Posts and Telecommunications (BUPT), Beijing, China. Her research interests include visible light communication and indoor positioning.
\end{IEEEbiography}

\vspace{-3ex}
\begin{IEEEbiography}[{\includegraphics[width=1in,height=1.25in,clip,keepaspectratio]{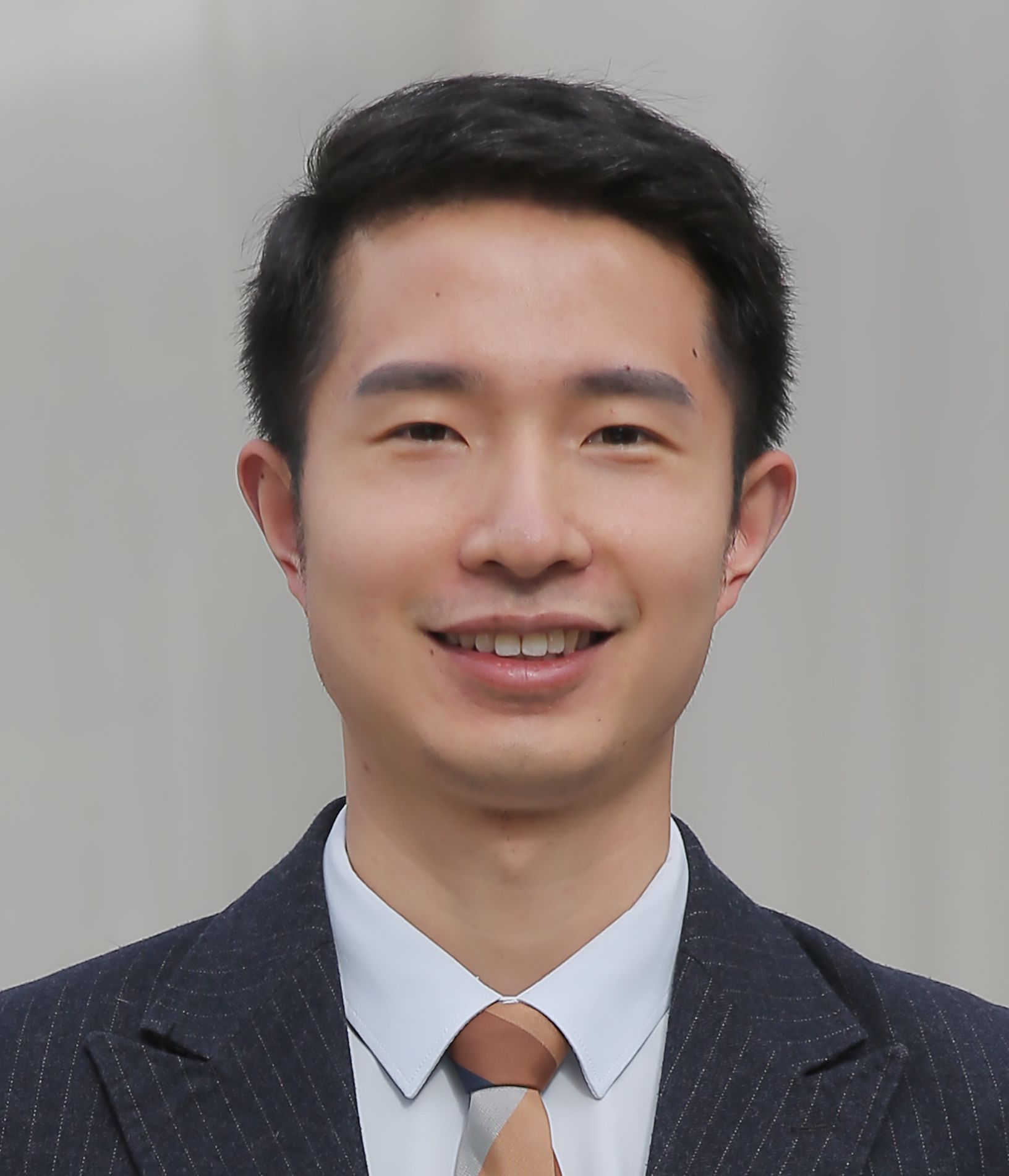}}]{Yang Yang} (Member, IEEE) is currently an Associate Professor with the School of Information and Communication Engineering, Beijing University of Posts and Telecommunications (BUPT), Beijing, China. He received the B.S. degree in information engineering from School of Communication, Xidian University, Xian, China, in June 2013, and the Ph.D degree in Information and Communication Engineering from School of Information and Communication Engineering, BUPT, Beijing, China.  His research interests include indoor positioning and intelligent wireless communications. He was a guest editor of IEEE JSAC. He served as a symposium co-chair of WCSP 2023, workshop co-chair of WCNC 2023, and TPC member for a series of IEEE conferences including Globecom and ICC. He was a recipient of the IEEE WCNC 2021 Best Paper Award. 
\end{IEEEbiography}

\vspace{-3ex}
\begin{IEEEbiography}[{\includegraphics[width=1in,height=1.25in,clip,keepaspectratio]{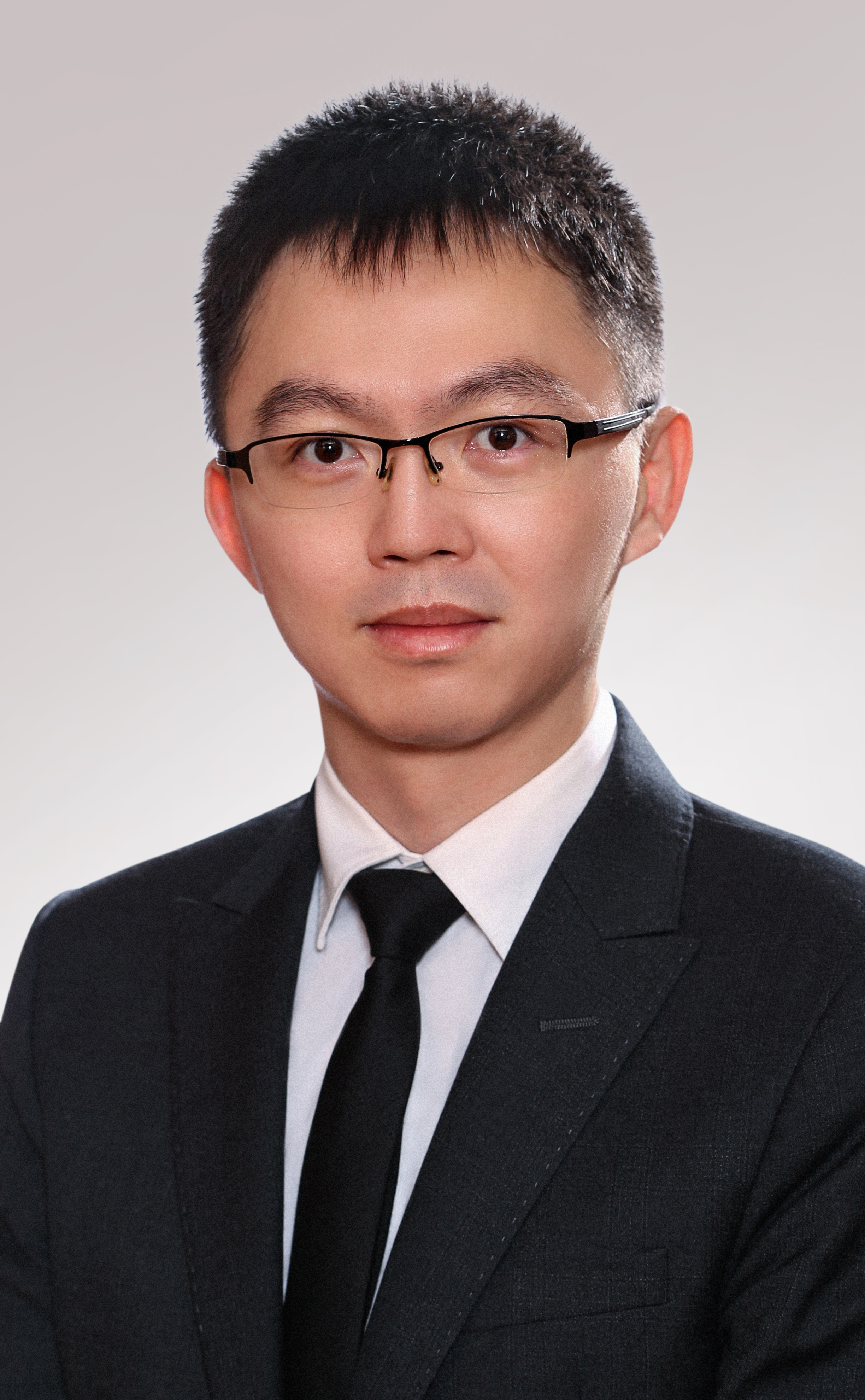}}]{Mingzhe Chen} (Senior Member, IEEE) is currently an Assistant Professor with the Department of Electrical and Computer Engineering and Frost Institute of Data Science and Computing at University of Miami. His research interests include federated learning, reinforcement learning, virtual reality, unmanned aerial vehicles, and Internet of Things. He has received four IEEE Communication Society journal paper awards including the IEEE Marconi Prize Paper Award in Wireless Communications in 2023, the Young Author Best Paper Award in 2021 and 2023, and the Fred W. Ellersick Prize Award in 2022, and four conference best paper awards at ICCCN in 2023, IEEE WCNC in 2021, IEEE ICC in 2020, and IEEE GLOBECOM in 2020. He currently serves as an Associate Editor of IEEE Transactions on Mobile Computing, IEEE Transactions on Communications, IEEE Wireless Communications Letters, IEEE Transactions on Green Communications and Networking, and IEEE Transactions on Machine Learning in Communications and Networking.
\end{IEEEbiography}

\vspace{-3ex}
\begin{IEEEbiography}[{\includegraphics[width=1in,height=1.25in,clip,keepaspectratio]{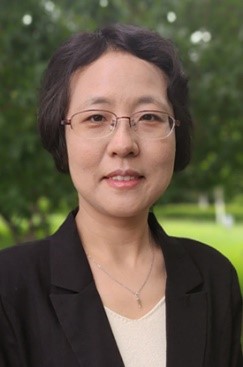}}]{Caili Guo} (Senior Member, IEEE) received the Ph.D. degree in Communication and Information Systems from Beijing University of Posts and Telecommunication (BUPT) in 2008. She is currently a Professor in the School of Information and Communication Engineering at BUPT. Her general research interests include machine learning and statistical signal processing for wireless communications, with current emphasis on semantic communications, deep learning, and intelligence-enabled edge computing for vehicle communications.
In the related areas, she has published over 200 papers and holds over 30 granted patents. She won Diamond Best Paper Award of IEEE ICME 2018 and Best Paper Award of IEEE WCNC 2021.
\end{IEEEbiography}

\vspace{-2cm}
\begin{IEEEbiography}[{\includegraphics[width=1in,height=1.25in,clip,keepaspectratio]{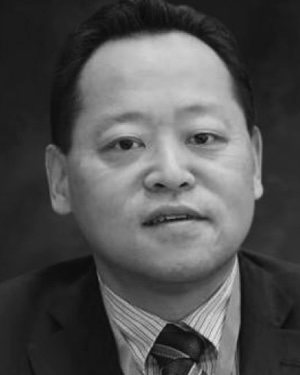}}]{Julian Cheng} (Fellow, IEEE) received the B.Eng. degree (Hons.) in electrical engineering from the University of Victoria, Victoria, BC, Canada, in 1995, the M.Sc. (Eng.) degree in mathematics and engineering from Queen’s University, Kingston, ON, Canada, in 1997, and the Ph.D. degree in electrical engineering from the University of Alberta, Edmonton, AB, Canada, in 2003. He was with Bell Northern Research, Kanata, ON, Canada; and NORTEL Networks, Ottawa, ON, Canada. He is currently a Full Professor with the School of Engineering, Faculty of Applied Science, The University of British Columbia, Kelowna, BC, Canada. His research interests include machine learning and deep learning for wireless communications, wireless optical technology, and quantum communications. Prof. Cheng was the Co-Chair of the 12th Canadian Workshop on Information Theory in 2011, the 28th Biennial Symposium on Communications in 2016, and the Sixth EAI International Conference on Game Theory for Networks in 2016. He was the General Co-Chair of the 2021 IEEE Communication Theory Workshop. He is the Chair of the Radio Communications Technical Committee of the IEEE Communications Society. He served as the President for the Canadian Society of Information Theory from 2017 to 2021. He was a past Associate Editor of IEEE Transactions on Communications, IEEE Transactions on Wireless Communications, IEEE Communications Letters, and IEEE Access, as well as an Area Editor of IEEE Transactions on Communications. He served as the Guest Editor for a Special Issue of IEEE Journal on Selected Areas in Communications on optical wireless communications. He is a registered Professional Engineer in BC, Canada.
\end{IEEEbiography}

\vspace{8cm}
\begin{IEEEbiography}[{\includegraphics[width=1in,height=1.25in,clip,keepaspectratio]{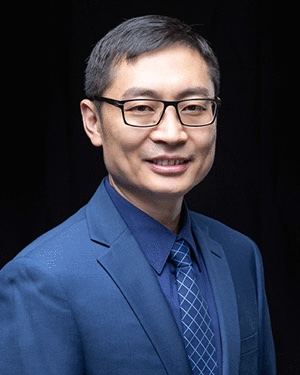}}]{Shuguang Cui} (Fellow, IEEE) received the PhD degree in electrical engineering from Stanford University, California, USA, in 2005. Afterwards, he has been working as assistant, associate, full, chair professor in electrical and computer engineering with the Univ. of Arizona, Texas A\&M University, UC Davis, and CUHK, Shenzhen respectively. He has also served as the executive dean for the School of Science and Engineering and is currently the director for Future Network of Intelligence Institute (FNii), CUHK, Shenzhen, and the executive vice director with the Shenzhen Research Institute of Big Data. His current research interests focus on data driven large-scale system control and resource management, large data set analysis, IoT system design, energy harvesting based communication system design, and cognitive network optimization. He was selected as the Thomson Reuters Highly Cited Researcher and listed in the Worlds’ Most Influential Scientific Minds by ScienceWatch, in 2014. He was the recipient of the IEEE Signal Processing Society 2012 Best Paper Award. He has served as the general co-chair and TPC co-chairs for many IEEE conferences. He has also been serving as the area editor for IEEE Signal Processing Magazine, and associate editors for IEEE Transactions on Big Data, IEEE Transactions on Signal Processing, IEEE JSAC Series on Green Communications and Networking, and IEEE Transactions on Wireless Communications. He has been the elected member for IEEE Signal Processing Society SPCOM Technical Committee (2009 2014) and the elected Chair for IEEE ComSoc Wireless Technical Committee (2017 2018). He is a member of the steering committee for IEEE Transactions on Big Data and the Chair of the steering Committee for IEEE Transactions on Cognitive Communications and Networking. He was also a member of the IEEE ComSoc Emerging Technology Committee. He was elected as an IEEE Fellow, in 2013, an IEEE ComSoc Distinguished Lecturer, in 2014, and IEEE VT Society Distinguished Lecturer, in 2019. In 2020, he won the IEEE ICC best paper award, ICIP best paper finalist, the IEEE Globecom best paper award. In 2021, he won the IEEE WCNC best paper award.
\end{IEEEbiography}
\end{document}